\documentclass[12pt]{article}
\pdfoutput=1
\usepackage{jheppub}
\usepackage{amsmath}
\usepackage{amsfonts}
\usepackage{amssymb}
\usepackage{graphicx}
\usepackage{float}

\usepackage[export]{adjustbox}
\newcommand{\bmat}{\left(\begin{array}}
\newcommand{\emat}{\end{array}\right)}

\def\yzero{\smash{\hbox{$y\kern-4pt\raise1pt\hbox{${}^\circ$}$}}}

\def\beq{\begin{equation}}
\def\eeq{\end{equation}}
\def\beqa{\begin{eqnarray}}
\def\eeqa{\end{eqnarray}}

\def\-{\hphantom{-}}

\def\s2{\frac{1}{\sqrt2}}

\def\beq{\begin{equation}}
\def\eeq{\end{equation}}
\def\beqa{\begin{eqnarray}}
\def\eeqa{\end{eqnarray}}

\def\IF{\relax{\rm I\kern-.18em F}}
\def\II{\relax{\rm I\kern-.18em I}}

\def\Dsl{\,\raise.15ex\hbox{/}\mkern-13.5mu D} 

\catcode`\@=11   
\newdimen\@rotdimen
\newbox\@rotbox  

\def\@vspec#1{\special{ps:#1}}
\def\@rotstart#1{\@vspec{gsave currentpoint currentpoint translate
   #1 neg exch neg exch translate}}
\def\@rotfinish{\@vspec{currentpoint grestore moveto}}
\def\@rotr#1{\@rotdimen=\ht#1\advance\@rotdimen by\dp#1%
   \hbox to\@rotdimen{\hskip\ht#1\vbox to\wd#1{\@rotstart{90 rotate}%
   \box#1\vss}\hss}\@rotfinish}
\def\@rotl#1{\@rotdimen=\ht#1\advance\@rotdimen by\dp#1%
   \hbox to\@rotdimen{\vbox to\wd#1{\vskip\wd#1\@rotstart{270 rotate}%
   \box#1\vss}\hss}\@rotfinish}%
\def\@rotu#1{\@rotdimen=\ht#1\advance\@rotdimen by\dp#1%
   \hbox to\wd#1{\hskip\wd#1\vbox to\@rotdimen{\vskip\@rotdimen
   \@rotstart{-1 dup scale}\box#1\vss}\hss}\@rotfinish}%
\def\@rotf#1{\hbox to\wd#1{\hskip\wd#1\@rotstart{-1 1 scale}%
   \box#1\hss}\@rotfinish}%
\def\rotate{\@ifnextchar[{\@rotate}{\@rotate[l]}}
\def\@rotate[#1]#2{\setbox\@rotbox=\hbox{#2}\@nameuse{@rot#1}\@rotbox}

\catcode`\@=12

\begin{document}

\makeatletter
\@addtoreset{equation}{section}
\makeatother
\renewcommand{\theequation}{\thesection.\arabic{equation}}
\pagestyle{empty}
\vspace*{0.5in}
\rightline{IFT-UAM/CSIC-24-62}
\vspace{1.5cm}
\begin{center}
\Large{\bf End of The World brane networks for infinite distance limits in CY moduli space
}
\\[8mm] 

\large{Roberta Angius\\[4mm]}
\footnotesize{Instituto de F\'{\i}sica Te\'orica IFT-UAM/CSIC,\\[-0.3em] 
C/ Nicol\'as Cabrera 13-15, 
Campus de Cantoblanco, 28049 Madrid, Spain}\\ 
\footnotesize{\href{roberta.angius@csic.es}{roberta.angius@csic.es}}

\vspace*{10mm}

\small{\bf Abstract} \\
\end{center}
\begin{center}
\begin{minipage}[h]{\textwidth}

Dynamical Cobordism provides a powerful method to probe infinite distance limits in moduli/field spaces parameterized by scalars constrained by generic potentials, employing configurations of codimension-1 end of the world (ETW) branes. These branes, characterized in terms of critical exponents, mark codimension-1 boundaries in the spacetime in correspondence of finite spacetime distance singularities at which the scalars diverge. Using these tools, we explore the network of infinite distance singularities in the complex structure moduli space of Calabi-Yau fourfolds compactifications in M-theory with a four-form flux turned on, which is described in terms of normal intersecting divisors classified by asymptotic Hodge theory. We provide spacetime realizations for these loci in terms of networks of intersecting codimension-1 ETW branes classified by specific critical exponents which encapsulate the relevant information of the asymptotic Hodge structure characterizing the corresponding divisors.           
\end{minipage}
\end{center}
\newpage
\setcounter{page}{1}
\pagestyle{plain}
\renewcommand{\thefootnote}{\arabic{footnote}}
\setcounter{footnote}{0}

\tableofcontents

\vspace*{1cm}

\newpage

\section{Introduction}

The study of compactifications in String Theory represents a powerful method for understanding the rich landscape of possible derived vacua and their corresponding phenomenology (see \cite{Polchinski:1998rr, Ibanez:2012zz, Blumenhagen:2013blt}). In this realm, the most interesting classes of compactifications are those that preserve only a small number of supersymmetries in the non-compact dimensions. Among these classes, a prominent one arises from compactifying M-theory on a Calabi-Yau fourfold manifold $Y_4$ \cite{Becker:1996gj,Dasgupta:1999ss, Haack:2001hl}, which leads to three-dimensional effective supergravity theory with $\mathcal{N}=2$ supersymmetry. Some of these vacua can be lifted to four-dimensional $\mathcal{N}=1$ F-vacua up to T-duality (see \cite{Heckman:2010hjj,Weigand:2010wti} for applications to phenomenology). Considering F-theory on $Y_4$, for specific choices of the compactification background, in particular when the manifold $Y_4$ admits an elliptic fibration with base $\mathcal{B}$, the limit in which the elliptic fiber shrinks to zero can be read as a four-dimensional vacuum obtained compactifying type IIB on the base $\mathcal{B}$. 

A central aspect in the study of Calabi-Yau compactifications controlled by several moduli is the intricate structure of their moduli space. These moduli parameterize geometric deformations of the compact manifold $Y_4$ controlling its total volume or the size of the internal cycles. It has been proven in \cite{Grimm:2018cpv} that infinite distance limits of this moduli space are singular loci corresponding to (de-)compactification regimes in the K\"ahler sector or points  in the complex structure sector where some internal cycle shrinks to zero size. For effective field theories with exact moduli space it is possible to explore these singular loci using spacetime independent scalar vevs. \\
 In the context of the swampland program \cite{Vafa:2005ui} (see also \cite{Palti:2019pca, vanBeest:2021lhn} for more recent reviews), many of the conjectures ruling out effective field theories that cannot be lifted into a consistent quantum gravity theory put constraints in the asymptotic behavior of these theories near the boundaries of their corresponding moduli space. This is the case of the Distance conjecture \cite{Ooguri:2006in}, which predicts a tower of exponentially light particles emerging when we approach one of these limits, and its sharpened versions \cite{Lee:2019wij,Etheredge:2022opl}, which provide information about the nature of these towers and set a lower bound on the decay rate of their masses. Similar constrains are imposed in terms of a Convex Hull condition \cite{Calderon-Infante:2020dhm} for the geodesic trajectories approaching the asymptotic regions of the field spaces of scalars constrained by non-trivial potentials. 
 
 In the presence of effective scalar potentials the adiabatic approach to investigate infinite distance limits with constant vevs is in general inconsistent \cite{Mininno:2011sdb} or forbidden \cite{Gonzalo:2019gjp} and the way to probe these limits is to make use of spacetime dependent solutions describing scalars that go to infinity in a finite distance in spacetime. These tools, first proposed in \cite{Buratti:2021yia,Buratti:2021fiv} (see also \cite{Etheredge:2022opl,Rudelius:2021azq,Calderon-Infante:2022nxb} for recent discussions on spacetime dependent solutions), have been subsequently defined within the framework of Dynamical Cobordisms, \cite{Angius:2022aeq,Angius:2022mgh,Blumenhagen:2022mqw,Angius:2023xtu,Blumenhagen:2023abk, Huertas:2023syg, Angius:2023rma} (see also \cite{Dudas:2000ff,Blumenhagen:2000dc,Dudas:2002dg,Dudas:2004nd} for early related works and \cite{Basile:2018irz,Antonelli:2019nar,Basile:2020xwi,Basile:2021mkd,Mourad:2021mas,Mourad:2021mas2,Mourad:2022loy,Basile:2022ypo} for more recent developments and \cite{Charmousis:2010cke,Kiritsis:2017knp} for holographic applications), describing configurations where the scalars run to infinity along a spacetime direction that ends at a finite distance in spacetime at which the spacetime metric features a Ricci singularity. These solutions can be regarded as networks of codimension-1 boundaries dressed by extended objects, i.e. End of The World (ETW) branes, sourcing the singularities and allowing spacetime to end. In this sense, Dynamical Cobordism solutions are especially interesting for the bottom-up exploration of these infinite distance limits. Some of these ETW branes do not have a UV resolution, or sometimes their corresponding miscorscopic description is unknown, see \cite{Friedrich:2024cob} for discussion in purely EFTs with a cutoff.
 
 Beyond these motivations, Dynamical Cobordism solutions represent the natural way to implement the Cobordism Conjecture \cite{McNamara:2019rup} in the framework of the effective field theories. In particular, they provide effective realizations of the End of The World configurations predicted by that conjecture at the topological level and hint at the presence of defects in the complete theory, realized as ETW branes in this effective approach, which are able to trivialize the corresponding cobordism group and make the compactification background bordant to nothing. 

 In the context of three dimensional supergravity theories, obtained by M-theory compactification on Calabi-Yau fourfolds, the effective potential admits a purely geometric description in terms of a $G_4$-flux turned on in some internal cycles of the compact manifold $Y_4$. The complex structure sector $\mathcal{M}_{cs}$ of the field space of these theories still holds a very intricate net of infinite distance singularities described in terms of normal crossing divisors \cite{Hironaka:1964} at which some internal four-cycle, maybe dressed with a $G_4$ flux, shrinks to zero size and produces a singular geometry for the corresponding $Y_4$. The mathematical formalism best suited to explore the structure of this network is encoded in the asymptotic Hodge theory \cite{Schmid:1973, Cattani:1986cks}. While in the bulk of the moduli space the middle cohomology $H^4 \left( Y_4, \mathbb{C} \right)$ admits a pure Hodge decomposition, as we approach a point in the boundary this structure is no longer valid and we need to define a finer structure, known as Deligne splitting, which includes the possibility to enhance four-forms to higher forms. The new splitting is completely determined by the nilpotent orbit approximation of the period vector and by the properties of local monodromy around the putative singular locus. This structure allows to provide a classification of the types of possible singular divisors forming the network and the allowed enhanced singularities occurring at their intersections \cite{Cattani:1986cks,kerr2019polarized}. Remarkably, the asymptotic structure of the middle cohomology also furnishes a suitable framework to formulate a growth theorem that is able to capture the leading growth of the Hodge norm of the four-forms near the boundary. The theorem allows to compute all the terms to construct the three-dimensional effective action controlling the dynamics of the scalars parameterizing the divisors involved in the putative local patch of the network. Applications of this mathematical machinery to the study of Calabi-Yau compactifications have been developed in several works\cite{Grimm:2019ixq,Grimm:2018cpv, Grimm:2018ohb, Grimm:2019grh, Bastian:2020bgh}, as well as applications to the computations of scattering amplitudes are starting to generate some interest \cite{Bonisch:2022, Vanhove:2014}.

 In this paper we will start to probe the network of infinite distance singularities of the complex structure sector of the moduli space associated with Calabi-Yau four-folds flux compactifications of M-theory using Dynamical Cobordism solutions of the corresponding three-dimensional effective action. We will present a dictionary associating to each singular divisor in $\mathcal{M}_{cs}$, classified in terms of its asymptotic Hodge-Deligne structure and equipped with the flux information, a specific codimension-1 ETW brane in spacetime, classified in terms of its critical exponent. Highly non-trivially, we will show that the consistency of the construction requires that the spacetime solution be time-dependent, and the  codimension-1 ETW brane mark a boundary for a timelike coordinate.
 
 In order to explore the intersections between distinct singular divisors in the moduli space, we will construct a new class of Dynamical Cobordism solutions involving intersecting ETW branes associated to two scalars attaining infinite field space distance in finite spacetime distance. The solution explores the infinite distance network of intersecting divisors, albeit in a subtle way, different from the naive expectation to associate to each intersecting ETW brane an intersecting divisor in the moduli space. The resulting spacetime picture nicely matches the moduli space view in \cite{Grimm:2018cpv, Grimm:2019ixq} that the intersection of divisors can be described as the enhancement of singularities in specific growth sectors
of the moduli along infinite distance paths. We will show that the structure of the scalar flux potentials has exactly the structure required to support the new spacetime dependent intersecting ETW brane Dynamical Cobordism solution.\\
 Although we are dealing with singular solutions in a regime where the effective field theory exhibits a lowered cutoff that limits its validity, these solutions describe the EFT version of objects that are well defined in the UV. This was also checked in a large classes of examples in \cite{Buratti:2021yia,Buratti:2021fiv,Angius:2022aeq} (see also \cite{Saracco:2012sat,Marchesano:2020rnd} for other setups in which singularities in supergravity are resolved by sources in the complete theory).  

 The paper is organized as follows. In section \ref{section2:generalities_CY_moduli} we review the main mathematical tools of the asymptotic Hodge theory necessary to characterize the network of infinite distance singularities in the complex structure sector of the Calabi-Yau moduli space. In section \ref{section3:DC} we will start reviewing the codimension-1 ETW brane solutions, following \cite{Angius:2022aeq}, and their intersecting configurations introduced in \cite{Angius:2023xtu}. In section \ref{section:decoupled_scalars} we will show how these intersecting configurations can be read in terms of decoupled scalar fields up to an appropriate redefinition of the fields.  In section \ref{section:Non-conformal-solutions} we will construct a new class of Dynamical Cobordism solutions involving two distinct divergent scalar fields and a non-conformally flat ansatz for the spacetime metric. These solutions can still be interpreted as intersecting configurations of two codimension-1 ETW branes up to an appropriate redefinition of the fields. In section \ref{section:4ETW_networks} we will exploit these Dynamical Cobordism solutions to probe the network of infinite distance singularities of $\mathcal{M}_{cs}$. In section \ref{section:ETW_brane_singularities} we will present the dictionary between singular divisors and ETW branes; in section \ref{section:ETW_branes_enhancements} we will apply the results of section \ref{section:Non-conformal-solutions} to the enhanced singularities occurring at the loci of intersections of singular divisors in $\mathcal{M}_{cs}$. Finally, we will provide some final considerations in section \ref{section:5_conclusions}.

\section{Generalities on Calabi Yau Moduli Space and Flux potentials}
\label{section2:generalities_CY_moduli}
In this introductory section we will review the mathematical tools necessary to provide a powerful local description of the complex structure moduli space of Calabi-Yau manifolds around its singular loci. Although the results are already in the literature, we review them to make our discussion self-contained, and to emphasize the key ideas relevant for the construction of our solutions in later sections. We keep the discussion brief, and refer the reader to \cite {Schmid:1973,Cattani:1986cks, Kashiwara:1985,kerr2019polarized} for more detailed explanations in the mathematical part and to \cite{Grimm:2018cpv,Grimm:2018ohb,Grimm:2019grh,Grimm:2019ixq, Grimm:2022gmh, Grimm:2023gjm} for the physical interpretation. The reader not interested in the mathematical details may take the results summarized in figure \ref{fig4} and table \ref{table_enhancements} and safely jump to section \ref{section3:DC}. 

A Calabi-Yau \textit{D-folds} manifold is a K\"ahler manifold of complex dimension $D$ admitting everywhere a non-vanishing $\left(D,0 \right)$ form $\Omega$. One way to fully determine a particular Calabi-Yau manifold $Y_D$ in a family of $CY_D$ manifolds is to specify its holomorphic form $\Omega$ and its K\"ahler $(1,1)-$form $J$. Deformations of these choices that keep the manifold in the same family can be of two different types: we can have \textit{complex structure} deformations parameterized by the choice of a $(D-1,1)$ form, or \textit{K\"ahler structure} deformations encoding all the possibilities to fix the K\"ahler form $J$. Geometrically, deformations of the first type control the size of the internal $D-$cycles, while deformations of the second type control the sizes of the even internal cycles of the Calabi-Yau. 

From now on we will consider the K\"ahler moduli to be fixed and we will focus our attention on the complex structure sector of the moduli space parameterized by deformations of $D-$cycles. For Calabi-Yau D-folds the complex structure manifold $\mathcal{M}_{cs}$ is still a K\"ahler manifold of complex dimension $h^{D-1,1}$, and it has the property of being neither smooth nor compact, \cite{Tian:1987} \cite{Todorov:1989}. This implies the existence of singular loci on it, whose corresponding Calabi-Yau manifolds are singular. Let us call this set of points $\Delta$ the \textit{discriminant locus} and assume that we can resolve it in terms of the union of normally intersecting divisors:
\begin{equation}
    \Delta = \bigcup_k \Delta_k.
    \label{discriminant_locus}
\end{equation}
Far away from these critical divisors, in the bulk of the moduli space,  the K\"ahler potential $\mathcal{K}^{cs} (z, \bar{z})$ is always a well-defined function:
\begin{equation}
    \mathcal{K}^{cs} (z, \bar{z})= - \log \int_{Y_D} \Omega (z) \wedge \ast \overline{\Omega} (\bar{z}) 
    \label{def:Kahler_potential}
\end{equation}
where the dependence on the moduli $\left\lbrace z^I \right\rbrace$, with $I=1,...,h^{D-1,1}$, is made explicit. This function induces a natural metric in $\mathcal{M}_{cs}$:
\begin{equation}
G_{I \overline{J}} = \partial_{z^I} \partial_{\overline{z}^J} \mathcal{K}^{cs}
\label{WP_metric}
\end{equation}
 known as Weil-Petersson metric.

 In the rest of the paper we will concentrate our attention on the study of $\mathcal{M}_{cs}$ associated with Calabi-Yau 4-folds. M-theory compactifications on this class of manifolds leads to three-dimensional supergravity with $\mathcal{N}=2$ supersymmetry where the complex structure moduli become the scalar fields for the effective action with kinetic terms induced by the metric \eqref{WP_metric}. An effective three-dimensional potential for these fields can be generated turning-on $G_4$ fluxes in the four-cycles of the internal $CY_4$ \cite{Becker:1996gj, Gukov:2000gvw, Dasgupta:1999ss, Giddings:2001yu}. Its definition in the three-dimensional Einstein frame is \cite{Haack:2001hl}:   
\begin{equation}
V_M= \frac{1}{\mathcal{V}_4^3} \left( \int_{Y_4} G_4 \wedge \star \overline{G}_4 - \int_{Y_4} G_4 \wedge G_4 \right).
\label{3d_scalar_potential}
\end{equation} 
The second term of this expression is topological and it is constrained by a \textit{tadpole cancellation} condition:
\begin{equation}
\frac{1}{2} \int_{Y_4} G_4 \wedge G_4 = \frac{\chi (Y_4)}{24},
\end{equation}
where $\chi (Y_4)$ is the Euler characteristic of $Y_4$. Assuming that this condition is satisfied, for the rest of the discussion we will limit our attention on the first contribution.\\
The expression \eqref{3d_scalar_potential} depends on both the complex structure and the K\"ahler moduli through the Hodge star operator and the volume $\mathcal{V}_4$ of $Y_4$. Since we are focusing our analysis in the complex structure sector, we impose the following condition on the flux $G_4$:
\begin{equation}
G_4 \wedge J =0,
\end{equation}
which limits our possibilities on the \textit{primitive cohomology} $H_p^4 \left( Y_4, \mathbb{R} \right)$\footnote{For the rest of the paper we restrict our attention on this primitive cohomology but we will omit the subindex $p$. } and isolates all  dependence on the K\"ahler moduli in the volume prefactor. 

\subsection{Variation of Hodge structure}

All the relevant quantities we introduced so far to construct the three-dimensional physical action depend on the complex structure through the holomorphic form $\Omega $ or the Hodge star operator in \eqref{3d_scalar_potential}. In this section we will review some essential mathematical tools that allow us to encode all these dependences in the language of variation of the Hodge structure.  

The nice property that the $k-th$ cohomology groups $H^k \left( Y_D, \mathbb{C} \right)$ of a smooth Calabi-Yau manifold admit a pure Hodge structure of weight $k$ means that for each level $k$ we can define a vector space $V_{\mathbb{C}}= H^k  \left( Y_D, \mathbb{C} \right)$, which always admits an Hodge decomposition:
  \begin{equation}
  V_{\mathbb{C}} = H^{k,0} \oplus H^{k-1,1} \oplus ... \oplus H^{1,k-1} \oplus H^{0,k} = \bigoplus_{k=p+q} H^{p,q} 
  \label{Hodge_decomposition}
  \end{equation}
  where the building subspaces satisfy the following complex-conjugation property:
  \begin{equation}
  H^{p,q} = \overline{H}^{q,p},
  \end{equation}
  and the weight $k$ is the constant sum of the indices $p+q=k$ of all the blocks in \eqref{Hodge_decomposition}. \\
  An equivalent way to rephrase the same property is saying that each cohomology group $H^k \left( Y_D, \mathbb{C} \right)$ defines a decreasing Hodge filtration:
   \begin{equation}
V_{\mathbb{C}} = F^0 \subset F^1 \subset ... \subset F^{k-1} \subset F^k = H^{k,0}
\end{equation}
where $H^{p,q} = F^p \cap \overline{F}^q$ and such that $F^p \oplus \overline{F}^{k+1-p} \cong H^k$.\\
If we can further equip our Hodge structure of a bilinear form $S(\cdot , \cdot)$ on $V_{\mathbb{C}}$ satisfying the following two properties:
\begin{equation}
\begin{split}
& (i) \quad \textsc{orthogonality} \quad S \left( H^{p,q} , H^{r,s} \right)=0 \quad \text{for} \quad p \neq s, q \neq r ; \\
& (ii) \quad \textsc{non-degeneracy} \quad i^{p-q} S \left( v, \overline{v} \right) >0 \quad \text{for} \quad v \in H^{p,q}, \quad v \neq 0 ,   \\
\end{split}
\end{equation}
the structure is called \textit{polarized}.

Since we are dealing with four-form in Calabi-Yau fourfolds we are interested in a deeper exploration of the middle cohomology $H^4 \left( Y_4, \mathbb{C} \right)$. In such a case the bilinear form is the cup product:
\begin{equation}
    S(v,w)= \int_{Y_4} v \wedge w \quad \quad \quad v,w \in H^4 \left(Y_4, \mathbb{C} \right),
\end{equation}
and it induces a norm for the vectors of the whole filtration:
\begin{equation}
\vert \vert v \vert \vert^2 = S(v,  \ast \overline{v})
\label{def:Hodge_norm}
\end{equation}
called Hodge norm.  

When we move on the moduli space the Hodge decomposition \eqref{Hodge_decomposition} changes. Roughly speaking due to change of what we call holomorphic and anti-holomorphic. One way to express this variation is in terms of the variation of the holomorphic four form $\Omega$ with respect to a fixed basis $\left\lbrace \gamma^{\mathcal{I}} \right\rbrace $ of $F^0$, with $\mathcal{I} = 1,..., dim H^4 \left( Y_4, \mathbb{C} \right)=2+2h^{3,1}+h^{2,2}$, such that the pairing:
\begin{equation}
    \eta_{\mathcal{I} \mathcal{J}} = - \int_{Y_4} \gamma^{\mathcal{I}} \wedge \gamma^{\mathcal{J}} 
\end{equation}
has signature $\left( 2h^{3,1} , 2+h^{2,2} \right)$.\\
In particular we can expand the form $\Omega$ along this basis writing:
\begin{equation}
\Omega = \Pi^{\mathcal{I}} (z) \gamma_{\mathcal{I}}.
\label{periods}
\end{equation}
The coefficients of such expansion are called \textit{periods} and they are in general complicated holomorphic trascendental functions of the moduli. Equivalently, they are defined by:
 \begin{equation}
 \Pi^{\mathcal{I}} (z) = \int_{\Gamma^{\mathcal{I}}} \Omega,
 \label{periods_def}
\end{equation}  
where $\Gamma^{\mathcal{I}}$ are the 4-cycles Poincaré-dual of the forms $\gamma^{\mathcal{I}}$. \\
At this point we can understand the variations of the spaces $F^p$ on $V_{\mathbb{C}}$ over the space $\mathcal{M}_{cs}$ in terms of the variations of the periods \eqref{periods_def} with respect to the coordinates $z$. 

This whole structure is completely well defined in the bulk of the moduli space, as long as the manifold $Y_4$ is smooth. However, when we approach a singular locus in $\mathcal{M}_{cs}$ the periods \eqref{periods_def} diverge, the Hodge filtration $F^p$ and the corresponding Hodge decomposition $H^{p,q}$ are not well defined anymore and we lose all information about what happens at these points.  This pushed the mathematical research to define analogous quantities at these limits, truncating the divergences of $F^p$ and $H^{p,q}$. 

\subsection{Asymptotic regime and N.O.T.}
\label{subsection:AR_NOT}

In this section we start exploring the boundaries of the complex structure moduli space $\mathcal{M}_{cs}$, in particular studying the asymptotic behavior of the Hodge structure of the middle cohomology within the nilpotent orbit approximation. 

\begin{figure}
    \centering
    \includegraphics[scale=0.12]{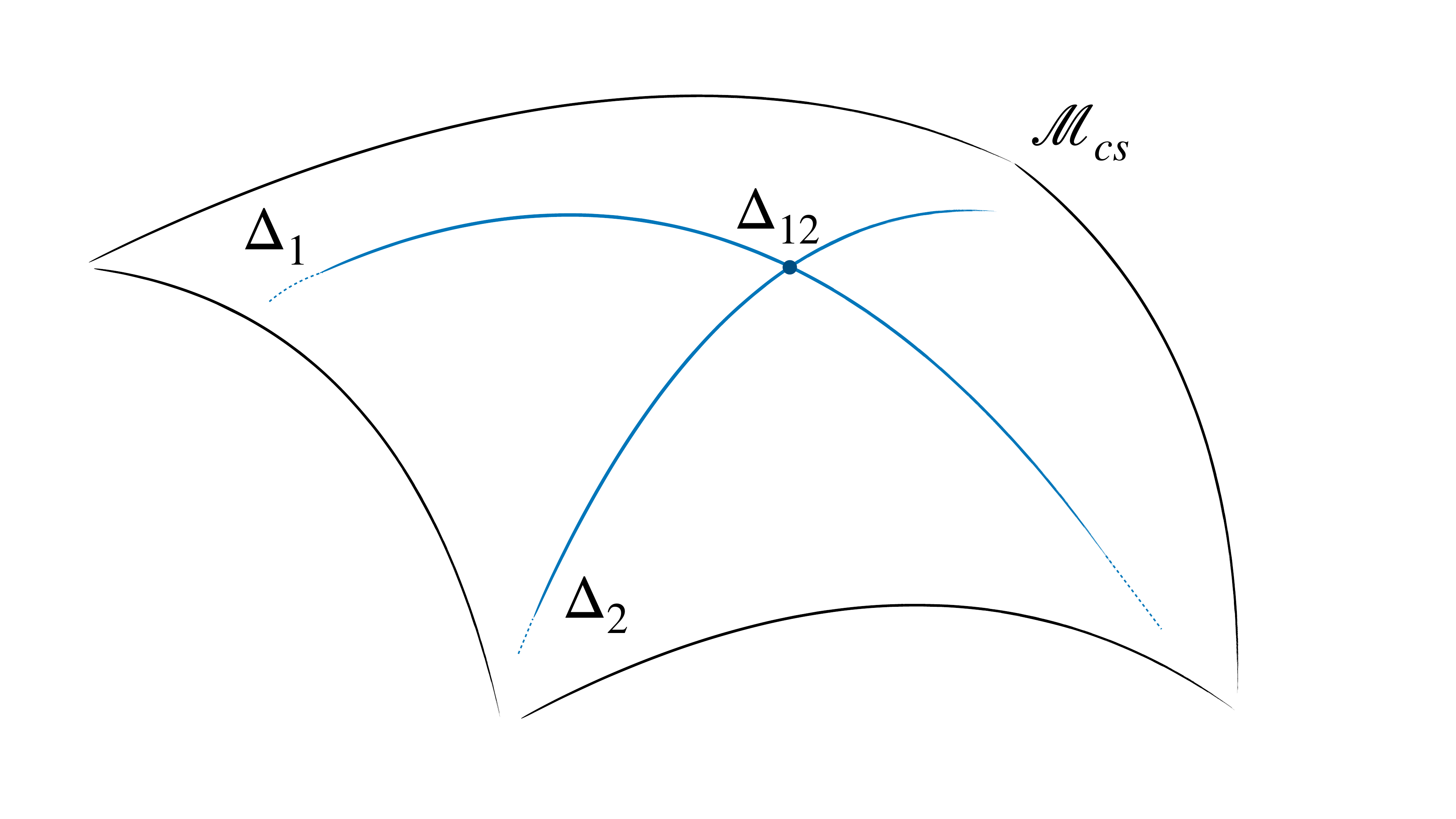}
    \caption{Intersection of two codimension-1 singular divisors $\Delta_1$ and $\Delta_2$ in the codimension-2 locus $\Delta_{12}$ of the two-dimensional complex moduli space $\mathcal{M}_{cs}$.}
    \label{fig1}
\end{figure}

As we pointed out above these asymptotic regions are resolved in terms of normal crossing divisors \eqref{discriminant_locus}, where each individual critical divisor $\Delta_k$ identifies a codimension-1\footnote{Note that when we talk about \textit{codimension} in the moduli space we mean a complex codimension. In later sections, when we will refer to spacetime codimension, it will mean a real codimension. } locus in $\mathcal{M}_{cs}$ and each normal intersection $\Delta_{k_1...k_n} = \Delta_{k_1} \cap ... \cap \Delta_{k_n}$ identifies a codimension-$n$ locus. See figure \ref{fig1} for an illustrative representation in a complex two-dimensional moduli space. \\
In order to understand what happens around these loci it is useful to introduce an adapted set of local coordinates $\left\lbrace z^i , \xi^{\bar{j}} \right\rbrace$  describing the patch $\mathcal{L} \subset \mathcal{M}_{cs}$ that contain the sub-discriminant locus $\Delta_{k_1 ... k_n}$ identified by the conditions $z^i=0$, with $i=1,..., n$ and $\bar{j}=n+1,...,h^{3,1}$. The structure of $\mathcal{L}$ is given by the product of $(h^{3,1}- n)$ disks $\mathcal{D}= \left\lbrace \xi \in \mathbb{C} \big\vert \vert \xi \vert < 1 \right\rbrace$, spanning the transverse directions to the singularity, and $n$ punctured disks $\mathcal{D}^{\ast}= \left\lbrace z \in \mathbb{C} \big\vert 0 < \vert z \vert < 1 \right\rbrace$, which span the longitudinal directions:
\begin{equation}
\mathcal{L} = \left( \mathcal{D}^{\ast} \right)^{n} \times \left( \mathcal{D} \right)^{h^{3,1}- n}.
\label{disks_topology}
\end{equation}
Using a conformal transformation we can map the punctured disks $\mathcal{D}^{\ast}$ in the upper half complex plane via:
\begin{equation}
    z^i \quad \mapsto \quad t^i= \frac{1}{2 \pi i} \log z^i = a^i + i s^i.
    \label{local_coordinates}
\end{equation}
With respect to these new suggestive coordinates the boundary $z^i=0$ is reached sending to infinity the imaginary part $s^i$ of $t^i$ and leaving the real part $a^i$ constant. Moreover, the periods are multivalued functions of the moduli, due to monodromies in the phase of the coordinates $z^i$, which in the new coordinates are thus associated to shifts of $a^i$. Let us encode the transformation properties of the periods around each singular divisor $\Delta_k$ in the \textit{monodromy matrices} $T_k$ acting conventionally as:
\begin{equation}
\mathbf{\Pi} \left( ..., e^{2 \pi i} z^k, ... \right) = T_k^{-1} \Pi \left( ..., z^k , ... \right).
\end{equation}
These local monodromies will be sufficient to classify all the types of singularities in $\mathcal{M}_{cs}$ and to extract information about the behavior of the Hodge structure of the middle cohomology around these limits. In particular, the crucial information for this analysis is contained in the infinite order part of these matrices, which can be extracted through the factorization:
\begin{equation}
T_k = T_k^{(s)} \cdot T_k^{(u)},
\end{equation}
where $T_k^{(s)} $ is the finite order part and $T_k^{(u)}$ is the \textit{unipotent} part. For the rest of the discussion we will only care about this last factor defining the following related matrix:
\begin{equation}
N_k = \log \left( T_k^{(u)} \right),
\end{equation} 
which is \textit{Nilpotent}, namely there exists a positive integer $n_k$ such that $N_k^{n_k+1}=0$. Matrices $N_k$ associated with different divisors $\Delta_k$
are commutative. 

In the asymptotic regime, reached in the limit $t_1,...,t_{n} \mapsto + \infty$, the \textit{Nilpotent Orbit Theorem} (N.O.T.) \cite{Schmid:1973} states that the period vector is represented by the following expansion:
\begin{equation}
\mathbf{\Pi} (t^i, \xi^{\bar{j}}) = e^{ \sum_{i=1}^n t^iN_i} \left( \mathbf{a}_0 (\xi)  + \mathbf{a}_j (\xi) e^{2 \pi i t^j} + \mathbf{a}_{ij} (\xi) e^{2 \pi i (t^i + t^j )} +...   \right),
\end{equation}
where the entries of the vectors $\mathbf{a}_{\bullet} (\xi)$ are holomorphic functions, in general non-polynomial, of the non-singular coordinates $\xi^{\bar{j}}$. This implies that, up to exponential corrections, the nilpotent orbit:
\begin{equation}
\mathbf{\Pi}_{nil} (t^i, \xi^{\bar{j}}) = e^{\sum_{i=1}^n t^iN_i}  \mathbf{a}_0 (\xi)
\end{equation}
is a good approximation of the period vector near the locus $\Delta_{k_1 ...k_n}$. \\
Following this expansion, we can define near each asymptotic region the \textit{limiting filtration}:
\begin{equation}
    F^p_{lim} \left( \Delta_{k_1 ... k_n} \right)= \lim_{t_{k_1},...,t_{k_n} \mapsto + i \infty} e^{- \sum_{i=1}^n t^i N_i} F^p (t),
\end{equation}
which is related with the pure Hodge structure $F^p$ by the N.O.T. and it has the property that is stays finite.

Just like the Hodge filtration $F^p$ produces the decomposition \eqref{Hodge_decomposition} of the middle cohomology, the limiting filtration $F^p_0$, together with the information about the monodromies $N_1, ..., N_n$, packaged in any element $N$ of the cone $\sigma (N_1,N_2,...,N_n) = \left\lbrace \sum_{i=1}^n a^i N_i \vert a^i > 0 \right\rbrace$, contains all the ingredients to construct a \textit{mixed} Hodge decomposition where the middle cohomology lifts into a finer splitting $\left\lbrace I^{p,q} \right\rbrace$, with $0 \leq p, q \leq 4$, known as \textit{Deligne splitting}:
\begin{equation}
H_p^4 \left( Y_4, \mathbb{C} \right) \quad \longrightarrow \quad \bigoplus_{0 \leq p, q \leq 4} I^{p,q}
\label{Deligne_splitting}
\end{equation}

The significant ingredient that allows us to give a formal definition for this splitting is encoded in the vector spaces:
\begin{equation}
W_i (N) =  \sum_{j \geq max (-1,l-4)}ker N^{j+1} \cap Im N^{j-l+4},
\end{equation}
producing a monodromy weight filtration under the action of $N$ such that \cite{Cattani:1982cka}: 
\begin{equation}
    N W_l \subseteq W_{l-2}
\end{equation}

Using the vector spaces $W_i$ and $F^p_{lim}$ we can define the Deligne splitting through the formula:
\begin{equation}
    I^{p,q} = F_{lim}^p \cap W_{p+q} \cap \left( \overline{F}^q_{lim} \cap W_{p+q} + \sum_{j \geq 1} \overline{F}^{q-j}_{lim} \cap W_{p+q-j-1}\right).
    \label{def:Deligne_splitting}
\end{equation}
This is the unique definition such that the following three properties are satisfied:
\begin{itemize}
\item[\textit{(i)}] $F^p_{lim} = \bigoplus_{r \geq p} \bigoplus_s I^{r,s}$;
\item[\textit{(ii)}] $W_l = \bigoplus_{p+q \leq l} I^{p,q}$;
\item[\textit{(iii)}] $I^{p,q} = \overline{I^{q,p}}$ mod $\bigoplus_{r < p , s < q} I^{r,s}$.
\end{itemize}
Acting on the space $I^{p,q}$ with the Nilpotent matrix $N$, we have:
\begin{equation}
    N I^{p,q} \subset I^{p-1,q-1}.
\end{equation}
However, in general not the whole lower $(p,q)-$ spaces can be obtained through this action. The subspaces $P^{p,q} \in I^{p,q}$ that cannot be obtained acting with $N^k$ on $I^{p+k,q+k}$ form the \textit{primitive part} of the splitting, and we have:
\begin{equation}
    I^{p,q} = \oplus_{i \geq 0} N^i \left( P^{p+i,q+i} \right),
    \label{I_primitive_decomposition}
\end{equation}
The elements of $P^{p,q}$  satisfy the following \textit{polarization conditions}:
\begin{equation}
\begin{split}
    & S \left( P^{p,q}, N^l P^{r,s} \right) =0 \quad \quad \quad \text{for} \quad p+q=r+s =l+4 \quad \text{and} \quad \left( p,q \right) \neq \left( s, r\right) \\
    & i^{p-q} S \left( v, N^{p+q-4} \overline{v} \right) > 0 \quad \quad \quad \text{for} \quad v \in P^{p,q} , \quad v \neq 0, \\
\end{split}
\end{equation}
which guarantees us that the elements belonging to the primitive part have strictly-positive norm. These conditions will become important to study the allowed Deligne splittings occurring at the enhanced singularities. It will be crucial that they are correctly transmitted, and this will put severe constraints on the form of the enhancement.

\subsection{Strict Asymptotic regime and $SL(2)-$orbit theorem}
\label{subsection:SAR}

In this work we are especially interested to construct the effective three-dimensional actions for the complex structure moduli near any asymptotic region of $\Delta$. This requires to have a prescription to approximate the Hodge norm of four-forms in order to explicitly compute the kinetic metric \eqref{WP_metric} and the leading behavior of the scalar potential \eqref{3d_scalar_potential}. The prescription is given by the \textit{Growth Hodge-norm theorem}, proven in \cite{Cattani:1986cks}, using the information of the monodromy matrices $N_i$ and the vector $\mathbf{a}_0$.\\
Another fundamental data we need when we treat with singular loci of codimension higher than 1 is the specification of the growth sector we use to reach the final singularity. This means to choose an order to send the moduli to infinity in the asymptotic region around $\Delta_{k_1...k_n}$. Each of these choice defines a growth sector:
\begin{equation}
\mathcal{R}_{12...n} = \left\lbrace t^i = a^i +i s^i \big\vert \frac{s^1}{s^2} > \gamma, \frac{s^2}{s^3} > \gamma, ... \frac{s^{n-1}}{s^n}> \gamma, s^n > \gamma , \varphi^i < \delta  \right\rbrace.
\label{growthsector}
\end{equation}
For $\gamma >>1$ this definition specifies what we call a \textit{strict asymptotic regime} (SAR) around $\Delta_{k_1...k_n}$.
For application to the growth theorem, the objects that better encode the previous data are a set of $n$ $sl(2, \mathbb{C})$ commuting algebras:
\begin{equation}
    sl \left( 2, \mathbb{C} \right)_i = \langle N_i^- , N_i^+, Y_i \rangle,
\end{equation}
associated to the $i-$th divisor $\Delta_{k_i}$ involved in the intersection $\Delta_{k_1...k_n}$. The triplets $\left( N_i^- , N_i^+, Y_i \right)$ can be computed following a recursive method applied to the splittings $\left\lbrace \tilde{I}^{p,q}_{(i)} \right\rbrace$ which are associated with all the loci $\Delta_{k_1...k_i}$, with $i<n$, traversed to reach the highest intersection $\Delta_{k_1...k_n}$ following the order of \eqref{growthsector}. These refined splittings are $\mathbb{R}-$split\footnote{They satisfy the property \textit{(iii)} below \eqref{Deligne_splitting} with zero modulo.} Deligne splitting computed using the filtration $\hat{F}^p$ related to $F^p$ up to the action of two operators $\hat{\xi}$ and $\hat{\delta}$:
\begin{equation}
    \hat{F}^p= e^{\hat{\xi}} e^{-i \hat{\delta}} F^p.
\end{equation}
The $Sl(2)-$\textit{orbit theorem} in \cite{Cattani:1986cks} prove this statement and provide an explicit construction of the matrices $\hat{\xi}$ and $\hat{\delta}$.\\
Now, starting from the higher Deligne splitting $\left\lbrace \tilde{I}^{p,q}_{(n)} \right\rbrace$ we can compute the operator $Y_{(n)}=Y_1 + ... + Y_n$ such that:
\begin{equation}
Y_{(n)} \tilde{I}_{(n)}^{p,q} (\Delta_{k_1...k_n}) = ( \underbrace{p+q}_{l_n} -4 ) \tilde{I}_{(n)}^{p,q}.
\end{equation}
Then, repeating the construction for all the subsequent splittings $\left\lbrace I^{p,q}_{(i)} \right\rbrace$ until the operator $Y_1$ associated to the divisor $\Delta_1$, we can reconstruct all the operators $Y_i = Y_{(i)}- Y_{(i-1)}$. At this point, decomposing $N_i$ into the basis of the eigenvectors of $Y_{(i-1)}$, namely $N_i = \sum_{\alpha} N_i^{\alpha}$, we can define $N_i^- := N_i^0$, which is the only element in the decomposition that commutes with $Y_{(i-1)}$. Finally, we can complete the triples looking for the elements $N^+_i$ satisfying the correct commutation relations and preserving the polarization. 

The first important consequence of the $Sl(2)$-orbit theorem is the definition of a further approximation of the nilpotent orbits in the strict asymptotic regime, specified by the sector $R_{12...n}$, with the $Sl(2)-$orbits:
\begin{equation}
\mathbf{\Pi}_{Sl(2)} = e^{i \sum_{i=1}^n s^i N^-_i} \tilde{\mathbf{a}}_0^{(n)} (\xi)
\label{def:a0}
\end{equation}
related with $\mathbf{\Pi}_{nil}$ via:
\begin{equation}
\mathbf{\Pi}_{nil} = e^{ \sum_{i=1}^n t^i N_i} \mathbf{a}_0 (\xi) = e^{ \sum_{i=1}^n a^i N_i} \cdot M(s) \cdot \mathbf{\Pi}_{Sl(2)},
\label{Pi_nil/Pi_Sl}
\end{equation}
where $M(s)$ is a $s^{i}-$dependent matrix introduced in section $5$ of \cite{Schmid:1973}.\\
This new approximation drops the subleading polynomial corrections in $\mathbf{\Pi}_{nil}$.  

The second relevant consequence for our aim is the fact that the $sl \left( 2, \mathbb{C}\right)$ triplets allow to decompose the cohomology group as:
\begin{equation}
    H^4_p \left( Y_4 , \mathbb{C} \right) = \bigoplus_{\mathbf{l} \in \mathcal{E}} V_{\mathbf{l}}
    \label{H4_decomposition}
\end{equation}
where the entries of the vector $\mathbf{l}= (l_1, l_2,...,l_n)$ are related to the eigenvalues of $\mathbf{v_l}$ with respect the operators $Y_{(i)}$ as:
\begin{equation}
Y_{(i)} v_{\mathbf{l}} =(l_i -4) v_{\mathbf{l}}.
\end{equation}
Note that this decomposition holds near the locus $\Delta_{k_1 ... k_n}$ and strictly depends on the growth sector we use to reach that locus. Now, any vector $w \in H_p^4 \left( Y_4, \mathbb{C} \right)$ can be written in an unique way as:
\begin{equation}
    w = \sum_{\mathbf{l} \in \mathcal{E}} w_{\mathbf{l}},
\end{equation}
where $w_{\mathbf{l}} \in V_{\mathbf{l}}$. For each of these vectors the growth Hodge-norm theorem ensures that the leading growing of its norm in the $Sl(2)-$approximation is captured by a term of the form:
\begin{equation}
\vert \vert w \vert \vert^2 \sim \textit{const} \cdot \left( \frac{s^1}{s^2} \right)^{l_1-4} ...  \left( \frac{s^{n-1}}{s^n} \right)^{l_{n-1}-4} (s^n)^{l_n-4}, \quad \quad \quad \textit{const} >0,
\label{growth_theorem}
\end{equation}
where the list of numbers $(l_1, ..., l_n)$ identifies the location of $w$ in the intersection of monodromy filtrations:
\begin{equation}
w \in W_{l_1} (N_1) \cap W_{l_2} (N_1+N_2) \cap ... \cap W_{l_n} (N_1 + N_2 +...+N_n).
\end{equation}
We will exploit this theorem to compute the kinetic metrics and the flux potentials for the three-dimensional spacetime action in section \ref{section:4ETW_networks}.

\subsection{Classification of singularities}
\label{section:singularities}

In this section we make use of the mathematical tools explained above in order to classify the singularities that can occur in the complex structure moduli space $\mathcal{M}_{cs}$ of Calabi-Yau 4-folds compactifications. This implies a classification of all the possible Deligne splittings \eqref{Deligne_splitting} associated with the points of $\Delta$. For the moment we consider to approach the discriminant locus from the bulk of the moduli space towards one of its singular divisors $\Delta_k$ in a region very far away from any higher intersection.

\begin{figure}[h!]
    \centering
    \includegraphics[scale=0.15]{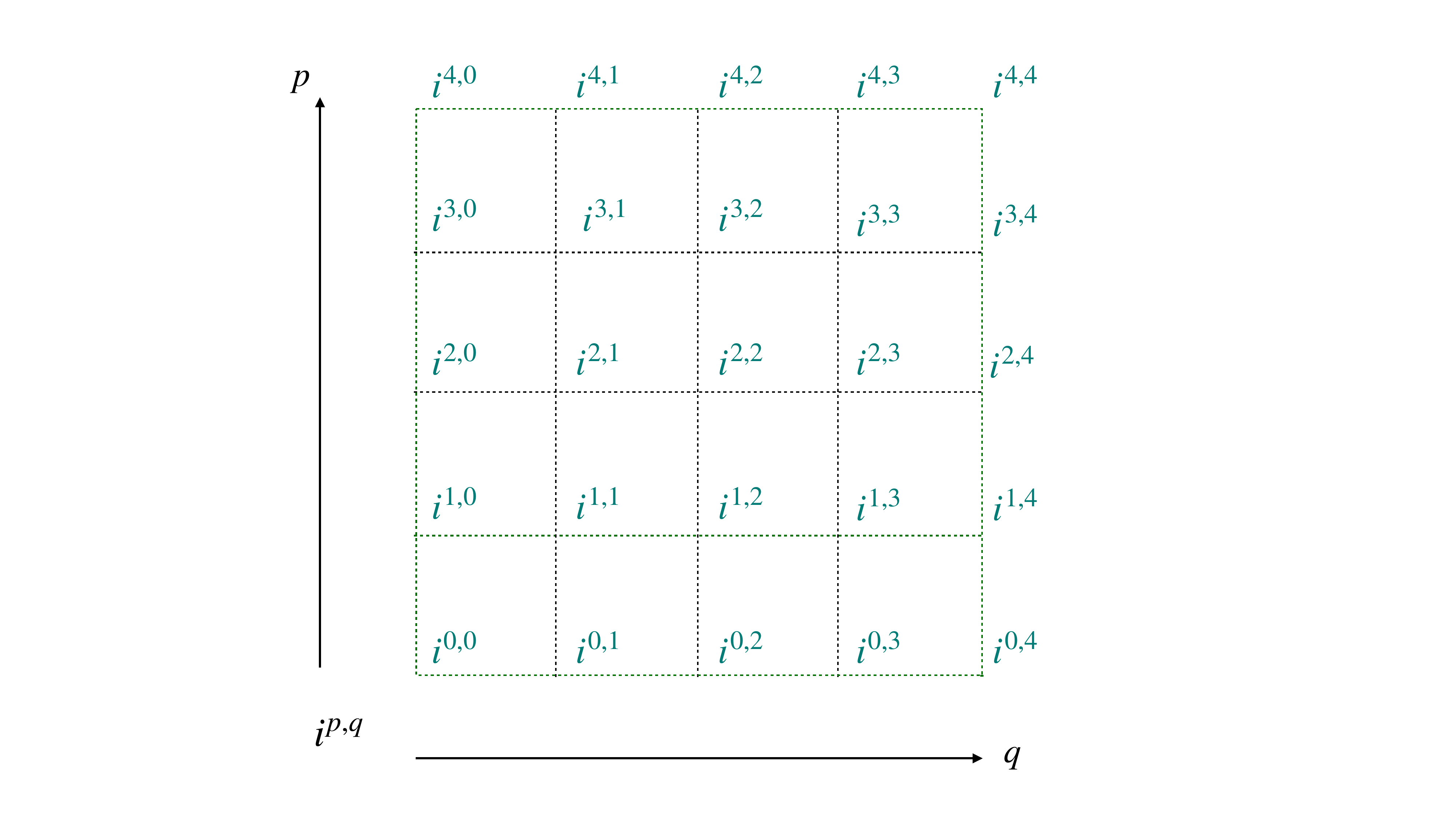}
    \caption{Hodge-Deligne diamond encoding the complex dimensions of the Deligne spaces $dim_{\mathbb{C}}I^{p,q}=i^{p,q}$}
    \label{fig2}
\end{figure}

In order to better visualize the discussion we associate to each splitting $\left\lbrace I^{p,q} \right\rbrace$ a lattice representing the associated Hodge-Deligne diamond, as in figure \ref{fig2}, where each lattice point $(p,q)$ is labeled with the dimension $i^{p,q}$ of the corresponding space $I^{p,q}$.

These complex dimensions satisfy the following properties:
\begin{equation*}
\begin{split}
    & \textit{(i)} \quad  i^{p,q} =i^{q,p} ; \quad \quad \quad  \quad \quad  \quad \textit{(iii)} \quad i^{p,q} \leq i^{p+1,q+1}, \text{for } p+q \leq 2 \\
    & \textit{(ii)} \quad i^{p,q} = i^{4-q,4-p}; \quad \quad \quad  \quad  \textit{(iv)} \quad \sum_{q=0}^4 i^{p,q} = h^{p,4-p}.\\
\end{split}
\end{equation*}
For Calabi-Yau 4-folds we have a unique holomorphic four form $\Omega$, then $h^{4,0}=h^{0,4}=1$. Moreover the property \textit{(iv)} tells us that we have only five possibilities to spread this value in the new Hodge-Deligne decomposition:
\begin{equation}
1=h^{4,0}= i^{4,d} \quad \quad \quad \quad \quad \text{with} \quad \quad \quad \quad  d= 0, 1, 2, 3, 4.
\end{equation}
According to \cite{kerr2019polarized}, these possibilities fix the Roman label of five large classes of singularities:
    \begin{align*}
        d \quad  = \quad   &0,&  &1,& &2,&  &3,&  &4& \\
        &\downarrow&  &\downarrow& &\downarrow& &\downarrow& &\downarrow& \\
        &I&  &II& &III& &IV& &V&  
    \end{align*}
    
\begin{figure}[h!]
    \centering
    \includegraphics[scale=0.22]{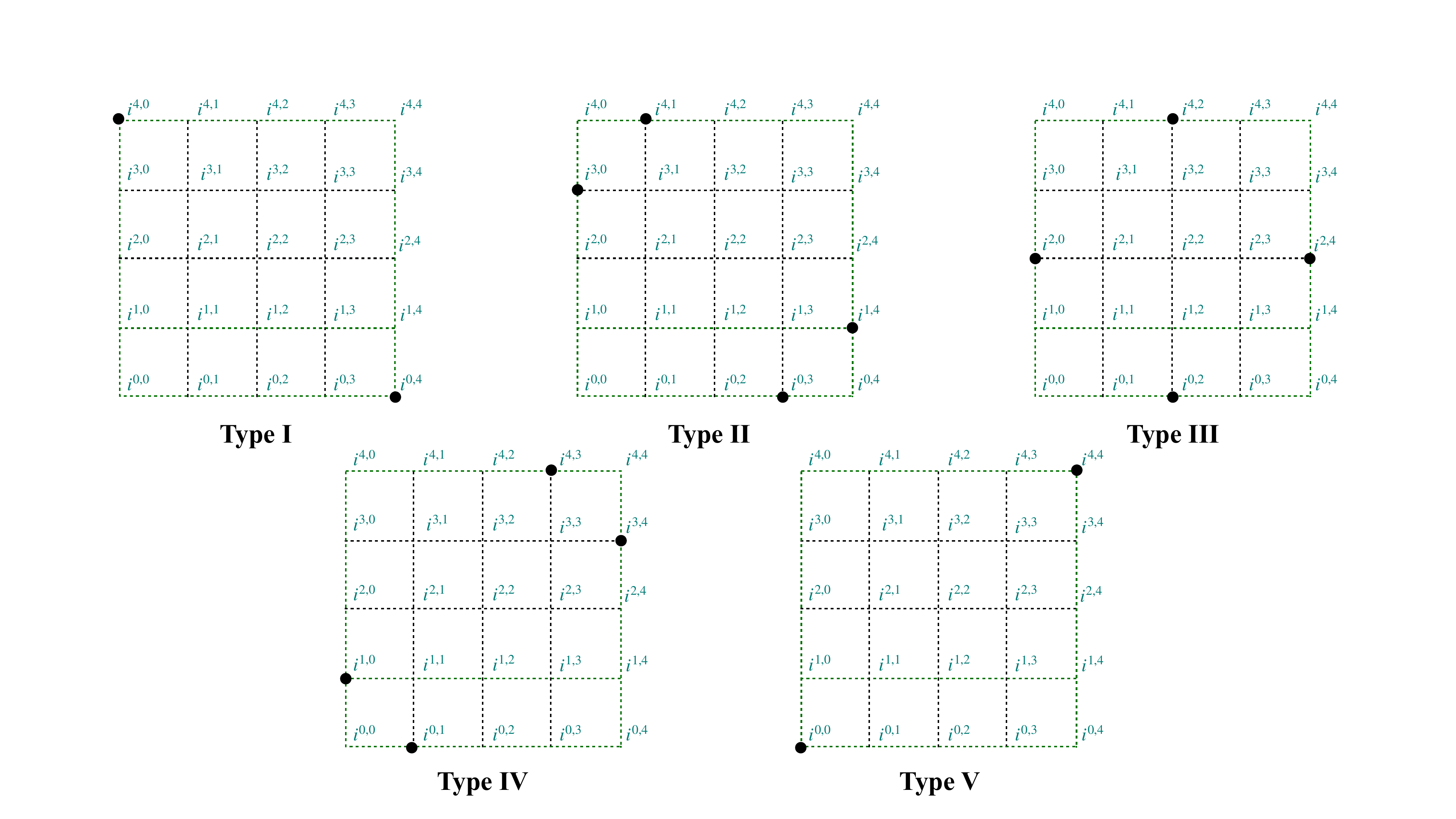}
    \caption{Basic lattices classifying the five large classes of singularities for Calabi-Yau 4-folds.}
    \label{fig3}
\end{figure}

Let us now introduce the notation to indicate the dimensions of the Deligne spaces with the number of dots on the corresponding lattice points. Using the properties \textit{(i)-(ii)} we can associate to each Roman class a basic lattice over which we can build the remaining unspecified components of the splitting. These basic lattices are summarized in figure \ref{fig3}. Note that the Roman label of the class uniquely determines the dots on the external perimeter of the diamond. In order to fully specify the type of singularity, we still need to fix the number of dots on the internal square.

Before proceeding, we need to specify the class of Calabi-Yau 4-folds manifold we are dealing with. In particular we have to fix the dimension $h^{1,3}=h^{3,1}$ of the groups $H^{1,3}$ and $H^{3,1}$.  Such information specifies the dimension of $\mathcal{M}_{cs}$ and, with the property \textit{(iv)}, it tells us that the total number of dots along the columns $(1,q), (3,q)$ and the rows $(p,1), (p,3)$ has to be $h^{1,3}$.  

To make the description more explicit, we will focus on the case $h^{1,3}=2$,  but the same procedure can be repeated analogously for other classes of Calabi-Yau manifolds. \\
Let us also fix $h^{2,2}=\widehat{m}$, so that the sum of dots along the central column and row in the diamonds must be $\widehat{m}$. 

\begin{figure}[H]
    \centering
    \includegraphics[scale=0.2]{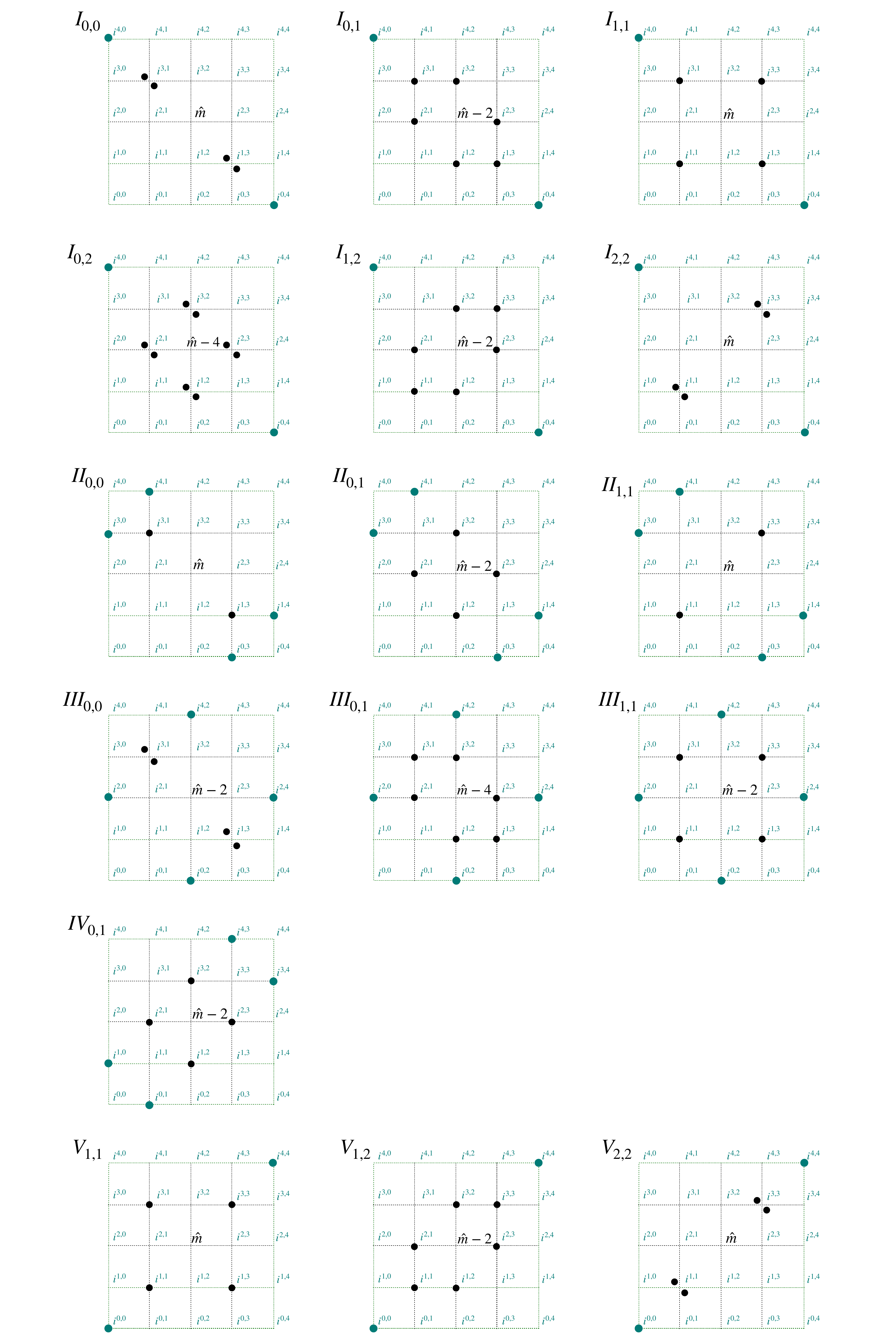}
    \caption{Hodge-Deligne diamonds associated with all the possible singularities of the $h^{3,1}=2$ dimensional complex structure moduli space of Calabi-Yau 4-folds with $h^{2,2}=\hat{m}$.}
    \label{fig4}
\end{figure}

Starting with the five models depicted in figure \ref{fig3}, and using the properties \textit{(i)-(iv)}, we can construct all the Hodge-Deligne diamonds representing all the possible singularities \cite{Grimm:2019ixq}. To distinguish between all of these possibilities we add two subindices $(a,a')$ to each Roman number: the first index indicates the dimension $i^{3,3}$ and the second one the sum $i^{3,3} +i^{3,2}$.  All allowed diamonds are summarized in the figure \ref{fig4}. In the next sections we will not consider type $I_{a,a'}$ singularities because they are located at finite distance in the moduli space as explained in section \ref{section:ETW_brane_singularities}.

\subsection{Allowed enhancements}
\label{section:enhancements}
In the previous section we classified all the allowed simple singularities that occur in $\mathcal{M}_{cs}$ approaching a singular divisor $\Delta_k$. Now we want to consider enhanced singularities which occur when two singular divisors intersect. In the case $h^{1,3}=2$ the intersection loci $\Delta_{k_1 k_2} = \Delta_{k_1} \cap \Delta_{k_2}$ are 0-dimensional loci (points). 

As explained in section \ref{subsection:SAR} we can reach the locus $\Delta_{k_1 k_2}$ following paths in $\mathcal{M}_{cs}$ belonging to different growth sectors. From now on we will focus our attention on the sectors $R_{12}$, but the same procedure can be implemented also for the sectors $R_{21}$.
\noindent
Starting from a generic point in the bulk of the moduli space, we send the first modulus $s^1 \mapsto  \infty$ to get the singularity $\textit{Type}_A \left( \Delta_{1} \right)$. Then we send also the second modulus $s^2 \mapsto \infty$ to reach the intersection $\Delta_{12}$ with the divisor $\Delta_{2}$. The singularity occurring at this point must be one of those classified in the previous section. Let us call such singularity $\textit{Type}_B \left( \Delta_{12} \right)$ and indicate the enhancement through the sector $R_{12}$ with the following notation:
\begin{equation}
\textit{Type}_A \left( \Delta_1 \right) \quad \quad \longrightarrow \quad \quad \textit{Type}_B \left( \Delta_{12} \right).
\label{enhancement}
\end{equation}

\begin{figure}[h!]
    \centering
    \includegraphics[scale=0.12]{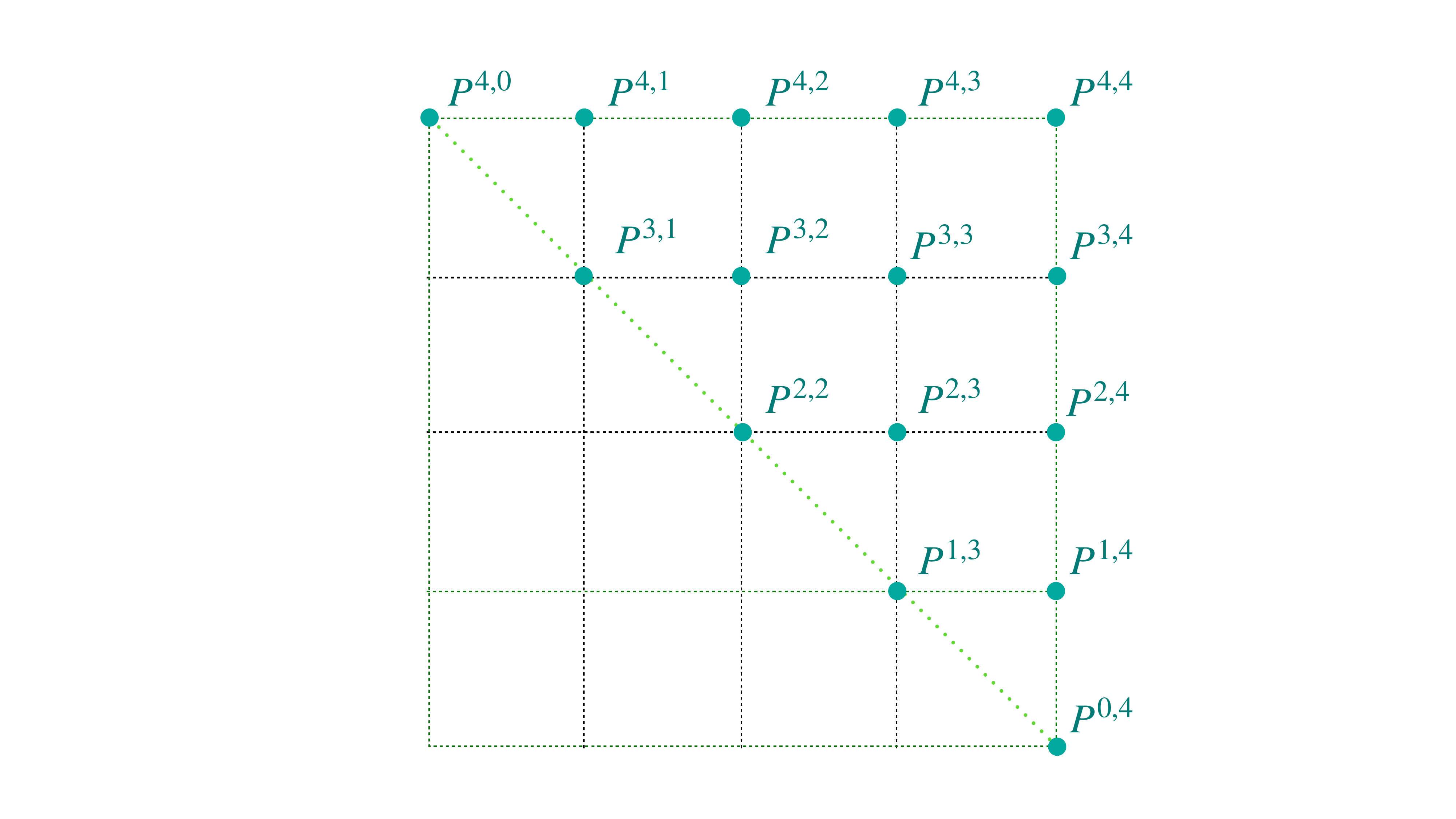}
    \caption{Principal parts of the Deligne splitting for the middle cohomology of $CY$ $4-$folds.}
    \label{fig5}
\end{figure}

Roughly speaking we have to understand which possible Deligne-splitting $ \left\lbrace I^{p,q} \left( \Delta_{12} \right) \right\rbrace$ can enhance from the Deligne splitting $\left\lbrace I^{p,q} \left( \Delta_{1} \right) \right\rbrace$. The fundamental mathematical object containing this information is the primitive part of  $\left\lbrace I^{p,q} \left( \Delta_{k_1} \right) \right\rbrace$, which can be written as:
\begin{equation}
P^{p,q} = I^{p,q} \cap \ker N_1^{p+q-3}.
\label{primitive_part}
\end{equation}
Using this definition and the equation \eqref{I_primitive_decomposition}, we make explicit in figure \ref{fig5} which spaces $I^{p,q}$ could contain a non-trivial primitive part. These primitive parts can be grouped in the following vector spaces:
\begin{equation}
\begin{split}
& P^8 = P^{4,4} \\
& P^{7} = P^{3,4} \oplus P^{4,3} \\
& P^{6} = P^{2,4} \oplus P^{3,3} \oplus P^{4,2} \\
& P^5 = P^{1,4} \oplus P^{2,3} \oplus P^{3,2} \oplus P^{4,1} \\
& P^4 = P^{0,4} \oplus P^{1,3} \oplus P^{2,2} \oplus P^{3,1} \oplus P^{4,0} \\
\end{split}
\end{equation}
each of which can be used to construct a \textit{pure Hodge structure} of respective weight $8,7,6,5,4$, exactly as $H^4_p$ admits a pure Hodge structure of weight $4$.
These new Hodge structures will be the basis to construct the next Deligne splitting $\left\lbrace I^{p,q} \left( \Delta_{12} \right) \right\rbrace$ occurring at the intersection. They will play the role of $H^4_p \left( Y_4, \mathbb{C} \right)$ for simple singularities.

For each $P^j$ we can construct a Deligne-splitting $\left\lbrace I^{p,q} \left( \Delta_{12} \right) \right\rbrace^j$, with $0 \leq p+q \leq 2j$, following the standard procedure described in section \ref{subsection:AR_NOT}. Then, there is a systematic way to rearrange these set of splittings to construct the Deligne-splitting $\left\lbrace I^{p,q} \left( \Delta_{12} \right) \right\rbrace$ occuring at the enhancement. This systematic way contains in itself the rules to determine whether a particular enhancement is admissible with respect to its capability to correctly transmit the polarization conditions.  We refer to \cite{kerr2019polarized} for a complete description of the procedure, while in the following we just summarize the result.  

Given an Hodge-Deligne diamond associated to a $j-$weight Hodge structure polarized by $S(\cdot, N \cdot)$, it is possible to define an integer-valued function on its corresponding lattice:
\begin{equation}
\diamondsuit \quad : \quad (p,q) \quad \longrightarrow \quad \mathbb{Z}
\end{equation}
such that the following properties are satisfied:
\begin{itemize}
\item[\textit{(i)}] $\sum_{q=0}^j \diamondsuit (p,q) = h^{p,j-p}$ for all $p$;
\item[\textit{(ii)}] $\diamondsuit (p,q) = \diamondsuit (q,p) = \diamondsuit (j-q,j-p)$ for all $p,q$;
\item[\textit{(iii)}] $\diamondsuit (p-1,q-1) \leq \diamondsuit (p,q)$ for $p+q \leq j$.
\end{itemize}
A simple choice for such a function is $\diamondsuit (p,q)=i^{p,q}$.
Using this definition we can easily define new Hodge-Deligne diamonds taking the sum of two diamonds:
\begin{equation}
\diamondsuit (p,q) = \diamondsuit_1 (p,q) + \diamondsuit_2 (p,q)
\end{equation}
or shifting the entries of one initial diamond:
\begin{equation}
\diamondsuit \left[ a \right] (p,q) = \diamondsuit (p+a,q+a).
\end{equation}
Now let $\diamondsuit \left( F_1, N_1 \right)$ be the Hodge-Deligne diamond associated with the singularity $\textit{Type}_A \left( \Delta_1 \right)$ and $\diamondsuit \left( F_2, N_2 \right)$ the one associated with the singularity $\textit{Type}_B \left( \Delta_{12} \right)$. From $\left( F_1, W (N_1) \right)$ we can construct the primitive vector spaces $P^j (N_1)$, and we can associate to each of these a $j-$weight pure Hodge structure polarized by $S(\cdot , N_1^j \cdot)$ with Hodge-Deligne diamond $\diamondsuit \left( F'_j, N^j_1 \right)$. \\
If the following decomposition is possible:
\begin{equation}
\diamondsuit \left( F_2 , N_2 \right) = \sum_{4 \leq j \leq 8} \sum_{0 \leq a \leq j-4} \diamondsuit \left( F'_j, N^j_1 \right) \left[ a \right]
\label{enhancement_rule}
\end{equation}
then the enhancement \eqref{enhancement} is allowed. 

In order to classify all the allowed enhancements in the space $\mathcal{M}_{cs}$ we have to check the condition \eqref{enhancement_rule} for all the enhancements we can construct using the singularities in figure \ref{fig4}. The results are summarized in table \ref{table_enhancements}, in agreement with table $(5.2)$ of \cite{Grimm:2019ixq}.

\begin{table}[H]
\centering
\begin{tabular}{|l|c|}
\hline
\textbf{Enhancement} & $(l_1 , l_2) \in \mathcal{E}$\\
\hline
$\textit{II}_{0,0}  \quad \longrightarrow \quad  \left\lbrace \begin{matrix} & \quad  \textit{II}_{0,1} \\ & \quad \textit{II}_{1,1} \end{matrix} \right. \quad $ & $\begin{matrix}  & (3,3), (4,3) , (4,4), (4,5), (5,5) \\ & (3,3), (4,2), (4,4), (4,6), (5,5)  \end{matrix}$ \\
\hline
$\textit{II}_{0,1} \quad \longrightarrow \quad \left\lbrace \begin{matrix} & \quad \textit{II}_{1,1} \\ & \quad \textit{III}_{0,0} \\ & \quad \textit{V}_{2,2} \end{matrix} \right.$ & $\begin{matrix} & (3,2), (3,3), (3,4), (4,4), (5,4), (5,5), (5,6) \\ & (3,2), (3,4), (4,4), (5,4), (5,6) \\ & (3,0), (3,2), (3,4), (3,6), (4,4), (5,2), (5,4), (5,6), (5,8) \end{matrix}$ \\
\hline
$\textit{III}_{0,0}  \quad \longrightarrow \quad \left\lbrace \begin{matrix} & \quad \textit{III}_{0,1} \\ & \quad \textit{III}_{1,1} \end{matrix} \right.$ & $\begin{matrix} & (2,2), (4,3), (4,4), (4,5), (6,6) \\ & (2,2), (4,2), (4,4), (4,6), (6,6)  \end{matrix}$ \\
\hline
$\textit{III}_{0,1} \quad \longrightarrow \quad \quad \quad \textit{III}_{1,1}$ & $(2,2),(3,2),(3,4),(4,4),(5,4),(5,6),(6,6)$ \\
\hline
$\textit{III}_{1,1} \quad \longrightarrow \quad \quad \quad \textit{V}_{2,2}$ & $(2,0),(2,2),(4,2),(2,4),(4,4),(6,4),(4,6),(6,6),(6,8)$ \\
\hline
$\textit{IV}_{0,1} \quad \longrightarrow \quad \quad \quad \textit{V}_{2,2}$ & $(1,0),(1,2),(3,2),(3,4),(4,4),(5,4),(5,6),(7,6),(7,8)$ \\
\hline
$\textit{V}_{1,1}  \quad \longrightarrow \quad  \left\lbrace \begin{matrix} & \quad \textit{V}_{1,2} \\ & \quad \textit{V}_{2,2} \end{matrix} \right.$ & $\begin{matrix} & (0,0),(2,2), (4,3), (4,4), (4,5), (6,6), (8,8) \\ & (0,0), (2,2), (4,2), (4,4), (4,6), (6,6), (8,8) \end{matrix}$ \\
\hline
$\textit{V}_{1,2} \quad \longrightarrow \quad \quad \quad \quad \textit{V}_{2,2}$ & $(0,0),(2,2),(3,2),(3,4),(4,4),(5,4),(5,6),(6,6),(8,8)$ \\
\hline
\end{tabular}
\caption{Allowed Enhancements for Calabi-Yau fourfolds with $h^{1,3}=2$.}
\label{table_enhancements}
\end{table}

Since we are considering flux compactifications and we need to compute the leading behavior of the potential \eqref{3d_scalar_potential} near each possible enhancement using the theorem \eqref{growth_theorem}, we have to know all the possible location of $G_4$ in the intersection of the monodromy filtrations:
\begin{equation}
    G_4 \in W_{l_1} \left(N_1 \right) \cap W_{l_2} \left( N_1 + N_2 \right).
\end{equation}
This is equivalent to classify all the possible doublets $(l_1 , l_2)$ that can contain some component of $G_4$. The table \ref{table_enhancements} summarizes the results for all available enhancements which corresponding expression of the flux potential will be explicitly studied in section \ref{section:ETW_branes_enhancements}. As mentioned before, we do not consider enhancements from type $I_{a,a'}$ singularities because they are located at finite distance in the moduli space.

\section{Dynamical Cobordisms}
\label{section3:DC}
Dynamical Cobordism techniques  \cite{Buratti:2021fiv,Buratti:2021yia,Angius:2022aeq,Angius:2022mgh,Angius:2023rma,Angius:2023xtu,Huertas:2023syg,Blumenhagen:2022mqw,Blumenhagen:2023abk} (see \cite{Dudas:2000ff,Blumenhagen:2000dc,Dudas:2002dg,Dudas:2004nd} for early related works and \cite{Basile:2018irz,Antonelli:2019nar,Basile:2020xwi,Basile:2021mkd,Mourad:2021mas,Mourad:2021mas2,Mourad:2022loy,Basile:2022ypo} for more recent developments, and \cite{Charmousis:2010cke,Kiritsis:2017knp} for holographic applications) represent a powerful method to explore large field regimes through dynamical solutions of generic spacetime actions. Moreover, in the context of the Swampland program, they are able to provide effective realizations of the configurations cobordant to nothing predicted at the topological level by the Cobordism Conjecture \cite{McNamara:2019rup}.     
\subsection{Codimension-1 ETW branes}
\label{Cod_1_ETW}
In this first section we provide a quick review of the method, that generalizes the construction in \cite{Angius:2022aeq} including also time-dependent solutions as discussed in \cite{Angius:2022mgh} (see also \cite{Dudas:2002dg,Basile:2018irz,Antonelli:2019nar,Mourad:2021mas} for early works on time-dependent solutions) in the simplest case of codimension-1 ETW branes. 

Consider the $d-$dimensional action in $M_P=1$ units:
\begin{equation}
    S= \int d^dx \sqrt{-g} \left[ \frac{1}{2} R - \frac{1}{2} \left( \partial \phi \right)^2 - V (\phi) \right]
    \label{action:single_scalar}
\end{equation}
containing Einstein-gravity coupled to a real scalar with arbitrary potential.\\
We consider solutions for the equations of motion associated to this action that run along one spacetime coordinate $y$ according with the ansatz:
\begin{equation}
    \begin{split}
        & ds^2_d = e^{-2 \sigma (y)} ds^2_{d-1} \pm dy^2 \\
        & \phi = \phi (y) , \\
    \end{split}
    \label{cod-1:ansatz}
\end{equation}
where the sign $\pm$ encompasses the possibilities to consider $y$ as a spacelike or a timelike coordinate. The condition for such a solution to realize a dynamical cobordism to nothing is the presence of a metric singularity at finite distance in spacetime corresponding to a divergent regime for the scalar. Imposing these requirements within the equations of motion, we obtain a very simple class of solutions with the following behavior for the fields:
\begin{equation}
\begin{split}
  & \phi (y) \simeq - \frac{2}{\delta} \log y \\
  & \sigma (y) \simeq - \frac{4}{\delta^2(d-2)} \log y. \\
\end{split}
\label{cod-1_solutions}
\end{equation}
The real number $\delta$ parameterizes the class of solutions and controls the growing of the scalar potential:
\begin{equation}
    V (\phi) = \mp a c e^{\delta \phi},
\end{equation}
where $c$ is a free parameter,  $a$ is related to $\delta$ by:
\begin{equation}
    \delta =2 \sqrt{\frac{d-1}{d-2}(1-a)}
\end{equation}
and the overall sign depends on whether $y$ is a spacelike or timelike coordinate. The equations of motion also impose the condition $a<1$ for this family, which imposes some bounds on $\delta$ when we choose a spacelike or a timelike running coordinate. In particular, if $V>0$ we have that, for $a<0$ (hence $\delta > 2 \sqrt{\frac{d-1}{d-2}}$) the coordinate $y$ is spacelike, and for $0<a<1$ (hence $\delta< 2 \sqrt{\frac{d-1}{d-2}}$) the coordinate $y$ is timelike; while if $V<0$, for $a<0$ (hence $\delta> 2 \sqrt{\frac{d-1}{d-2}}$) the coordinate $y$ is timelike, and for $0<a<1$ (hence $\delta< 2 \sqrt{\frac{d-1}{d-2}}$) the coordinate $y$ is spacelike.

Another interesting feature of these solutions is the existence of universal scaling relations linking the spacetime and the field space distances $\Delta$ and $\mathcal{D}$ with the spacetime scalar curvature $\vert R \vert$ in the following way:
\begin{equation}
    \Delta \sim e^{- \frac{\delta}{2} \mathcal{D}} , \quad \quad \vert R \vert \sim e^{\delta \mathcal{D}}.
    \label{scaling_relations_cod1_ETW}
\end{equation}
Such relations, controlled by the \textit{critical exponent} $\delta$, encode the defining properties for the realization of a dynamical cobordism in the spacetime solutions.

\subsection{Intersecting ETW branes}
\label{section:Intersecting_ETW}
In this section we overview the local description of intersecting ETW branes configuration done in \cite{Angius:2023rma}. These solutions, given by a mere superposition of branes, represent the simplest building block to probe multiple infinite distance limits in the field space. However, as we will discuss in section \ref{section:ETW_branes_enhancements}, these solutions are not adequate to probe infinite distance limits involving multiple scalars in the Calabi-Yau moduli space, but it will be necessary a generalization of the ansatz as described in section \ref{section:Non-conformal-solutions}.

We consider the following $(n+2)-$dimensional action containing Einstein gravity coupled to two real scalar fields constrained by the general potential $V \left( \phi_1 , \phi_2 \right)$:
\begin{equation}
    S= \int d^{n+2}x \sqrt{-g} \left\lbrace \frac{1}{2} R - \frac{1}{2} \left( \partial \phi_1 \right)^2 - \frac{1}{2} \left( \partial \phi_2 \right)^2 - \frac{\alpha}{2} \partial_{\rho} \phi_1 \partial^{\rho} \phi_2- V (\phi_1 , \phi_2) \right\rbrace.
    \label{action_2scalars}
\end{equation}
Note that the presence of a mixed term in the kinetic sector of the action is essential to solve the equations of motion for the class of solutions considered in \cite{Angius:2023rma}.

We consider the following structure for the spacetime metric:
\begin{equation}
ds_{n+2}^2\,=\, e^{-2\sigma_1-2\sigma_2}ds_n^2 + e^{-2\sigma_2}  dy_1^2 \pm  e^{-2\sigma_1}dy_2^2  \, ,
\label{metric-ansatz-codim2}
\end{equation}
which, by setting only one of the two running coordinates $y_i$ to a constant, nicely reduces to a local codimension-1 ETW metric up to a constant. This is equivalent to restricting our investigation to an $(n+1)$ dimensional $y_i-$constant slice in the $(n+2) $dimensional spacetime. The sign $\pm$ in the metric indicates the possibilities to consider the coordinate $y_2$ as a spacelike or a timelike direction. \\
Moreover, we require the following coordinate dependence for the scalars:
\begin{equation}
    \phi_1 = \phi_1 (y_1)\quad, \quad \quad \quad \phi_2 = \phi_2 (y_2).
\end{equation}

\begin{figure}[h!]
    \centering
    \includegraphics[scale=0.15]{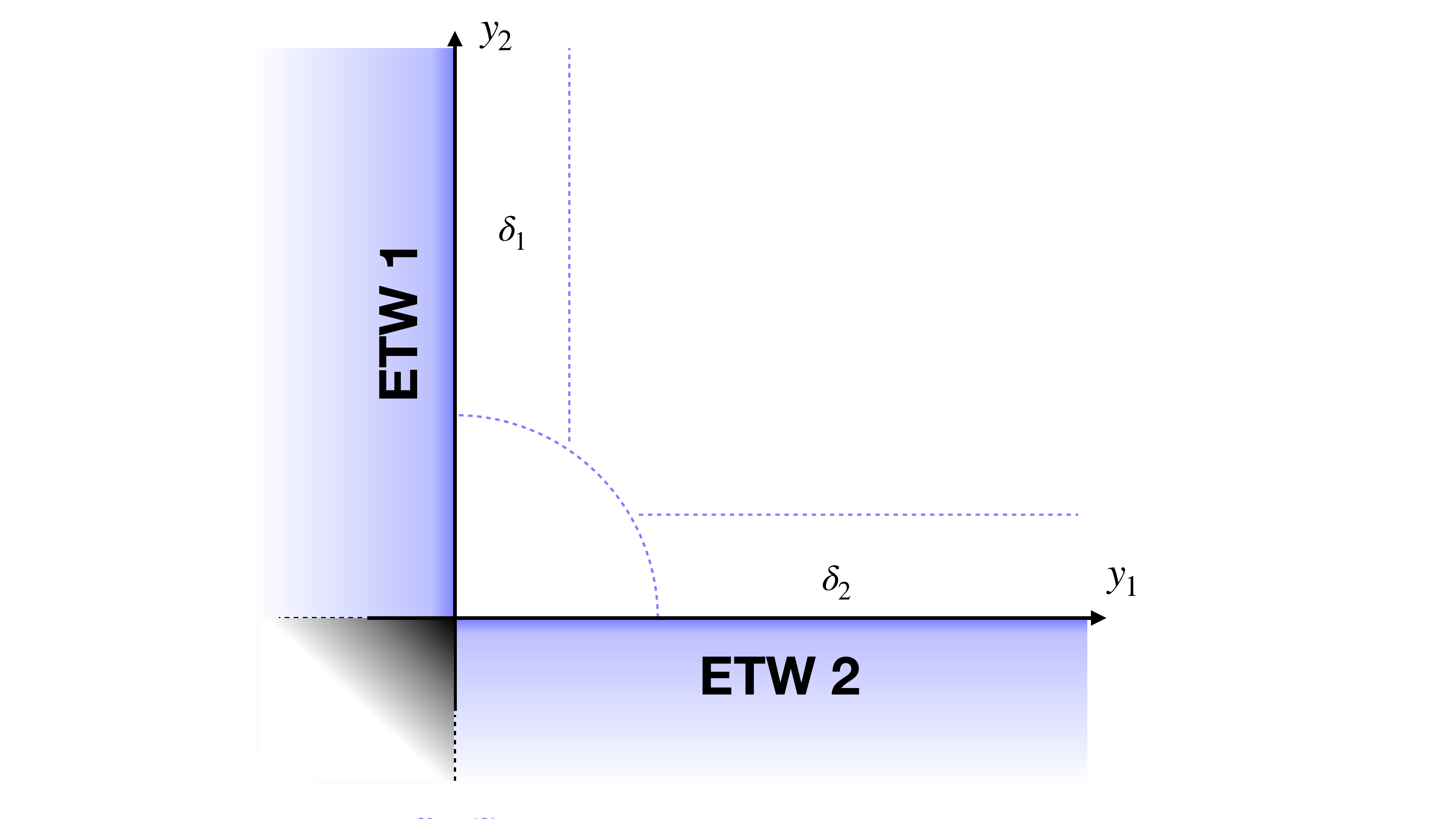}
    \caption{\small Spacetime configuration representing  two intersecting ETW branes of type $\delta_1$ and $\delta_2$, as figure 2 of \cite{Angius:2023rma}.}
    \label{fig6}
\end{figure}

A picture of the spacetime configuration described in this structure is shown in Figure \ref{fig6}, with two codimension-1 ETW branes intersecting at the codimension-2 locus $y_i=0$. 

Imposing the equations of motion, the local description near the codimension-2 locus is given by logarithmic functions:
\begin{equation}
\begin{split}
\phi_1&=-\frac{2}{\delta_1}\log y_1\quad   ,\quad \phi_2=-\frac{2}{\delta_2}\log y_2  \\
\sigma_1&= -\frac{4}{n\delta_1^2} \log y_1\quad ,\quad \sigma_2=-\frac{4}{n\delta_2^2}\log y_2 \\
\end{split}
\label{local-intersecting-etws}
\end{equation}
 where the parameters $\delta_i$ specify the type of intersecting branes and control the growing of the potential in the quarter circle area of figure \ref{fig6}:
 \begin{equation}
     V=V_1+V_2= - c_1v_1 \, e^{\delta_1\phi_1}e^{\frac{4}{n \delta_2} \phi_2}
 \mp c_2v_2\,e^{\frac{4}{n \delta_1}\phi_1}e^{\delta_2\phi_2}
 \end{equation}
 with $c_i$ positive integration constants.  The sign $\mp$ in the second term of $V$ refers to the cases where $y_2$ is a spacelike or a timelike coordinate, respectively.  The coefficients $v_i$ are related to the parameters $\delta_i$ as:
 \begin{equation}
     \delta_i^2 = \frac{8 (n+1)}{n + \sqrt{n \left[ n+8v_i (n+1) \right]}}.
     \label{critical-exp-intersecting-etws}
 \end{equation}

 The equations of motion also constrain the coefficient $\alpha$, controlling the mixed kinetic term in the action, to a specific value:
 \begin{equation}
     \alpha =  \frac{8}{n \delta_1 \delta_2}.
     \label{alpha}
 \end{equation}
 This highlights the fact that the solutions apply to a preferential basis of the field space which metric contain a non-trivial mixed term weighed by the parameter \eqref{alpha}. As we will explain in section \ref{section:ETW_branes_enhancements}, this invalidates their naive application in the Calabi-Yau moduli space, where each Calabi-Yau modulus is a scalar field, because in the corresponding infinite distance loci the asymptotic metrics are diagonal.  
 
\subsection{Intersecting configurations for decoupled scalar fields}
\label{section:decoupled_scalars}
 Since the Calabi-Yau moduli space metrics are diagonal in the asymptotic limit, as we will discuss in detail in section \ref{section:ETW_branes_enhancements}, we consider the following $(n+2)-$dimensional action for gravity coupled with two decoupled real scalar fields constraint by the general potential $V (\psi_1 , \psi_2)$:
\begin{equation}
    S = \int d^{n+2}x \sqrt{-g} \left\lbrace \frac{1}{2} R - \frac{1}{2} \left( \partial \psi_1 \right)^2 - \frac{1}{2} \left( \partial \psi_2\right)^2 - V(\psi_1, \psi_2)  \right\rbrace
    \label{action_2decoupled_scalars}
\end{equation}
Note that in this new action the dynamics of the scalars is controlled by a diagonal metric for the kinetic term. 

To solve the equations of motion, we consider the following ansatz for the fields: 
\begin{equation}
    ds^2_{n+2} =e^{2 A(y_1 , y_2)} ds^2_n + e^{2 B(y_1 ,y_2)} dy_1^2 \pm e^{2 C(y_1 , y_2)} dy^2_2
    \label{spacetime_metric:orthogonal}
\end{equation}
\begin{equation}
    \psi_1 = \psi_1 (y_1,y_2) \quad , \quad \quad \psi_2 = \psi_2 ( y_2).
\end{equation}
Note that, in order to find intersecting ETW brane solutions for scalars with diagonal kinetic terms,  we now allow one of the scalars to depend on two coordinates (although the equations of motion make this dependence relatively simple). Despite this could seem like a complication, we will actually show that it leads to an interpretation very much in the spirit of enhancements of singularities in CY moduli space in section \ref{section:ETW_branes_enhancements}.
A linearly independent set of combination of equations of motion for these functions is:

\begin{equation}
    \begin{split}
     (1) \quad & \mp e^{-2 C }n \left[ A' \left(n A'+B'-C' \right) + A'' \right]  - e^{-2B} \left[\dot{A} \left( n \dot{A}- \dot{B} + \dot{C} \right) + \ddot{A} \right] =2V \\
     (2) \quad &  e^{-2 B}  \left[ -n \dot{ A}^2 +(n-2) \dot{A} \left( \dot{C} - \dot{B} \right) +(n-2) \ddot{A}+2 \dot{C} \left( \dot{C} - \dot{B} \right) + 2 \ddot{C} + \left( \dot{\psi_1} \right)^2 \right] + \\
     \pm  e^{-2C} & \left[ -n A'' +(n-2) A' \left( B' -C' \right) +(n-2) A''+2 B' \left( B'- C' \right) +2 B'' + \left( \psi_1' \right)^2+ \left( \psi_2' \right)^2 \right] =0 \\
     (3) \quad & n \left[ \dot{A} B'- \dot{ A} A' + \dot{ C} B' - \dot{ A}' \right] = \dot{\psi_1} \psi_1' \\
     (4) \quad &  e^{-2 B} \left[ \left( n \dot{A} - \dot{ B} + \dot{C} \right) \dot{\psi_1} + \ddot{\psi_1} \right] \pm e^{-2C} \left[ \left( nA' +B'-C' \right) \psi_1' + \psi_1'' \right] - \frac{\partial V}{\partial \psi_1} =0 \\
     (5) \quad & \pm e^{-2 C} \left[ \left( n A' +B'-C' \right) \psi_2' + \psi_2'' \right] - \frac{\partial V}{ \partial \psi_2} =0, \\
    \end{split}
    \label{eoms_decoupled_scalars}
\end{equation}
where we have used the notation $\dot{f} := \partial_{y_1} f$ and $f' := \partial_{y_2} f$. The upper signs refer to the case where $y_2$ is a spacelike coordinate and the bottom signs to the case where $y_2$ is a timelike coordinate. \\
As in the previous section, we are interested in studying classes of solutions for these equations realizing a dynamical cobordism, with the scalars running to infinity in a finite distance in spacetime. We can conventionally choose that this happens as we approach the origin of the coordinates $y_i$.

Let us focus our attention in the simple class of solutions featuring an additive structure for the warp factors of the metric:
\begin{equation}
    A (y_1 , y_2) = - \sigma_1 (y_1) - \sigma_{2} (y_2) \quad , \quad B(y_1 , y_2) = - \sigma_2 (y_2) \quad , \quad C(y_1 , y_2) = - \sigma_1 (y_1)
    \label{conf_flat_ansatz}
\end{equation}
and logarithmic profiles for the functions $\sigma_i$ and $\psi_i$:
\begin{equation}
    \begin{split}
        \sigma_1 =- a_1 \log y_1 + \frac{1}{2} \log c_1 \quad &, \quad \sigma_2 = -a_2 \log y_2 + \frac{1}{2} \log c_2 \\
        \psi_1 =b_{11} \log y_1 +b_{12} \log y_2 \quad &, \quad \psi_2 =  b_{2} \log y_2 \\
    \end{split}
\end{equation}
Note that this means that at $y_1=0$ the scalar $\psi_1$ diverges, whereas at $y_2=0$ both scalars diverge. This will provide a nice match with the picture of singularity enhancement in CY moduli space of section \ref{section:enhancements}, as we explain later on. The functions $B$ and $C$ do not have a respectively $y_1$ and $y_2$ dependence because such a dependence can be easily reabsorbed by a change of variables in the metric. The real numbers $c_1$ and $c_2$ are integration constants, giving only subleading contributions in the $y_i\to 0$ limits.\\
Replacing these profiles in \eqref{eoms_decoupled_scalars} we get:
\begin{equation}
    \begin{split}
        & V= - \frac{1}{2} c_1 n a_1 \left( n a_1+ a_1-1 \right) y_1^{-2} y_2^{-2 a_2} \mp \frac{1}{2} c_2 n a_2 \left(n a_2 + a_2 -1 \right) y_1^{-2 a_1} y_2^{-2} = \\
        & := - c_1 v_1 y_1^{-2} y_2^{-2 a_2} \mp c_2 v_2 y_1^{-2 a_1} y_2^{-2}  \\
        & b_{11}^2 =na_1 \\
        & b_{12}^2 = na_1a_2^2 \\
        & b_2^2 = na_2 \left( 1- a_1 a_2 \right) \\
    \end{split}
\end{equation}
For $a_1, a_2 \neq 1$ the scalar potential splits in two pieces with a different dependence on $y_1$ and $y_2$, which means a different dependence on $\psi_1$ and $\psi_2$. These terms encode the leading contribution of the function $V$ in the asymptotic regime near the locus $y_i=0$. The constant coefficients $v_i$ are related to the parameters $a_i$ by:
\begin{equation}
      a_i= \frac{1 \pm  \sqrt{1 +8v_i \left( 1 + \frac{1}{n} \right)}}{2 (n+1)} 
\end{equation}

In order to give to the reader a clear overview on the results, we summarize below the spacetime solutions written in terms of the parameters $a_i$:
\begin{equation}
    \begin{split}
       & ds^2_{n+2} = y_1^{2 a_1} y_2^{2 a_2} ds^2_n + y_2^{2 a_2} dy_1^2 \pm y_1^{2 a_1} dy_2^2 \\
       & \psi_1 = - \sqrt{na_1} \log y_1 - a_2 \sqrt{na_1} \log y_2 \\
       & \psi_2 = - \sqrt{na_2 (1-a_1 a_2)} \log y_2. \\
    \end{split}
    \label{Solutions:class1}
\end{equation}
The potential takes the form:
\begin{equation}
    V = - c_1 v_1 e^{\frac{2}{\sqrt{n a_1}} \psi_1}  \mp c_2 v_2 e^{2 \sqrt{\frac{a_1}{n}} \psi_1} e^{2 \sqrt{\frac{1-a_1 a_2}{na_2}} \psi_2}.
    \label{potential:solution1}
\end{equation}
Note that the spacetime metric in \eqref{Solutions:class1} is conformally flat: this is a straightforward generalization of the metric \eqref{cod-1:ansatz} of the codimension-1 case. 

Actually, although we have constructed the above solutions from scratch from the action \eqref{action_2decoupled_scalars}, they can be regarded as a reinterpretation of the solutions in section \ref{section:Intersecting_ETW}. This is because the two solutions are related by a simple redefinition of the fields:
\begin{equation}
    \begin{split}
        & \phi_1 = \psi_1 - \left( \frac{a_1 a_2}{1-a_1 a_2} \right)^{1/2} \psi_2 \\
        & \phi_2 = \left( \frac{1}{1-a_1a_2} \right)^{1/2} \psi_2 \\
    \end{split}
    \label{field_redefinition}
\end{equation}
for which the spacetime action \eqref{action_2decoupled_scalars} takes the form \eqref{action_2scalars} with 
\begin{equation}
    \alpha = 2 \sqrt{a_1 a_2}. 
\end{equation}
Respect to these new fields we can read the solutions as the superposition of two codimension 1 ETW branes located at the respective positions $y_1=0$ and $y_2=0$, as depicted in figure \ref{fig6}, and characterized by the critical exponents $\delta_i$ given in \eqref{critical-exp-intersecting-etws}, which relation with the parameters $a_i$ is:
\begin{equation}
    \delta_i^2 = \frac{4}{n a_i}.
    \label{critical_exps:ai}
\end{equation}

Note that when we approach the locus $y_1 =0$ the new redefined scalar $\phi_1$ runs to infinity such as the scalar $\psi_1$ of the decoupled setup, while when we approach the locus $y_2=0$ only the scalar $\phi_2$ goes to infinity in the new fields redefinition but both the scalars $\psi_1$ and $\psi_2$ diverge at the same time in the decoupled picture. In section \ref{section:ETW_branes_enhancements} we will take advantage of this feature to associate to the ETW brane located at the position $y_1 =0$, where only the scalar $\psi_1$ diverges, a codimension-1 divisor in the Calabi-Yau moduli space, and to the other ETW brane, the one located at the position $y_2=0$ where both the scalars $\psi_1$ and $\psi_2$ diverge, not a second divisor intersecting with the first one but the intersection of the two divisors itself, i.e. the enhancement. 

Although this is a nice interpretation to solve the problem to have diagonal asymptotic metrics, the structure of the potential that support this class of solutions is not of the same form of the potentials that we get in the Calabi-Yau moduli space near the enhanced singularities, as we will see in section \ref{section:ETW_branes_enhancements}. In the next section we will introduce a generalization of the above solution, which is similar in spirit in the interpretation of the ETW branes, but it is supported by scalar potentials having the same form of the flux potentials computed using the growth theorem for the Hodge norm applied to the $G_4$ fluxes supported by the asymptotic Hodge structures allowed at the enhancements and summarized in table \ref{table_enhancements}.

\subsection{A new solution beyond the conformal flatness}
\label{section:Non-conformal-solutions}

In this section we present a generalization of the solutions described in the previous section and associated with the same action \eqref{action_2decoupled_scalars}. These new solutions are governed by a scalar potential of the right form to match the flux potential supported in the asymptotic regions of the Calabi-Yau moduli space around the enhancements. The simple modification with respect to the ansatz \eqref{conf_flat_ansatz} that we will consider in this section consists to go beyond the conformal flatness of the metric.  

We consider a non-conformally flat ansatz for the spacetime metric \eqref{spacetime_metric:orthogonal}:
\begin{equation}
\begin{split}
    & A (y_1, y_2) = a_1 \log y_1 + a_2 \log y_2 \\
    & B(y_1 , y_2)= a_2 \log y_2 - \frac{1}{2} \log c_2 \\
    &C (y_1 , y_2) = (1-a_1n) \log y_1 - \frac{1}{2} \log c_1 \\
\end{split}
\end{equation}
and we keep the same logarithmic profiles for the scalar fields:
\begin{equation}
    \psi_1 = b_{11} \log y_1 + b_{12} \log y_2 \quad , \quad \psi_2 = b_2 \log y_2.
\end{equation}
 The profile of the scalars is the same as in the previous section, so these solutions also follow the interpretation that one ETW brane corresponds to a singular divisor and the other ETW brane corresponds to the intersection of divisors, i.e. to the enhancement of the singularity.
 
Replacing these functions in the equations of motion \eqref{eoms_decoupled_scalars} we obtain the following constraints for the parameters:
\begin{equation}
    \begin{split}
        & b_{11}= - \sqrt{a_1 n(2-a_1-a_1n)} \\
        & b_{12} = - a_2 (1-a_1n) \sqrt{\frac{n}{a_1(2-a_1-a_1n)}}  \\
        & b_2 = - \sqrt{a_2 n \left[1+ \frac{a_2 (a_1n -1)^2}{a_1 (-2+a_1+a_1n)} \right]}\\
    \end{split}
    \label{solution2:ortho}
\end{equation}
and the following shape for the potential:
\begin{equation}
    V =  \mp \frac{c_2}{2} a_2 n \left( a_2 n +a_2 -1 \right) y_1^{-2+2a_1n} y_2^{-2} = \mp c_2 v_2 y_1^{-2+2a_1n} y_2^{-2}.
\end{equation}
As in the previous section, the upper signs refer to the case where $y_2$ is a spacelike coordinate and the bottom ones to the case where $y_2$ represents a timelike direction. Note that if $y_2$ is timelike, and $a_2 > \frac{1}{1+n}$, the potential is positive. For positive $a_i$ the relations \eqref{solution2:ortho} are well defined only if $a_1 < \frac{2}{1+n}$. 

For a clearer overall picture of the result, let us summarize the solutions by making the dependence on the spacetime coordinates explicit:
\begin{equation}
\begin{split}
    & ds^2_{n+2} = y_1^{2a_1} y_2^{2a_2} ds^2_n + y_2^{2a_2}dy_1^2 -y_1^{2(1-a_1n)}dy_2^2\\
    & \psi_1 = - \sqrt{a_1 n(2-a_1-a_1n)} \log y_1 -a_2 (1-a_1n) \sqrt{\frac{n}{a_1(2-a_1-a_1n)}} \log y_2 \\
    & \psi_2 = -\sqrt{a_2 n \left[1+ \frac{a_2 (a_1n -1)^2}{a_1 (-2+a_1+a_1n)} \right]}\log y_2. \\
\end{split}
\label{Solutions:class2}
\end{equation}
The potential written in terms of these fields takes the form:
\begin{equation}
    V = \mp c_2 v_2 e^{\frac{2-2a_1n}{\sqrt{a_1n (2-a_1-a_1n)}} \psi_1} e^{2 \sqrt{\frac{a_2+a_1 (-2+a_1+a_1n-2a_2n+a_1a_2n^2)}{a_1 a_2 n (-2+a_1+a_1n)}} \psi_2}.
    \label{potential:solution2}
\end{equation}
Note that with respect to the class of solutions treated in section \ref{section:decoupled_scalars}, here we have only a single term in the potential driving the dynamics of the fields near the locus $y_i=0$. This feature makes these solutions appropriate for the application to the infinite distance network of the Calabi-Yau moduli space. In fact, the structure \eqref{potential:solution2} of the potential is precisely the one that matches the flux potentials obtained from the Hodge norm of $G_4$ for all the admissible enhancements in the space $\mathcal{M}_{cs}$ with $h^{3,1}=2$.

In order to make more manifest the interpretation of this solution to describe the intersection of two codimension-1 ETW branes of the kind in section \ref{Cod_1_ETW}, we make use of the following redefinition of the fields:
\begin{equation}
    \begin{split}
        & \phi_1 = \psi_1 - (1-a_1n) \sqrt{\frac{a_2}{a_1(2-a_1-a_1n)-a_2(a_1n-1)^2}} \psi_2 \\
        &\phi_2 = \left( 1 + \frac{a_2 (a_1n-1)^2}{a_1(-2+a_1+a_1n)} \right)^{-1/2} \psi_2 \\
    \end{split}
    \label{redefined_fields}
\end{equation}
to write the spacetime action \eqref{action_2decoupled_scalars} in the form \eqref{action_2scalars}, with
\begin{equation}
    \alpha = 2(1-a_1n) \sqrt{\frac{a_2}{a_1(2-a_1-a_1n)}}
    \label{alpha_non_conf_flat}
\end{equation}
and  
\begin{equation}
    V = \mp c_2 v_2 e^{\frac{2-2a_1n}{\sqrt{a_1n (2-a_1-a_1n)}} \phi_1} e^{\frac{2}{\sqrt{a_2 n}} \phi_2},
\end{equation}

Importantly, one has to remember that the fields $\phi_1$ and $\phi_2$ should not be regarded as independent Calabi-Yau moduli, but rather that $\phi_1$ is a Calabi-Yau modulus and $\phi_2$ is a combination of $\phi_1$ and another orthogonal modulus. In this interpretation, the non-trivial mixed term \eqref{alpha_non_conf_flat} emphasizes the fact that $\phi_2$ is not orthogonal to $\phi_1$.

With respect to these new fields we can read the solutions as the intersection of two codimension 1 ETW branes located at the respective positions $y_1 = 0$ and $y_2 = 0$. And we characterize these branes using the critical exponents:
\begin{equation}
    \begin{split}
        &\delta_1 = \frac{2-2a_1n}{\sqrt{a_1n (2-a_1-a_1n)}} \\
        & \delta_2 = \frac{2}{\sqrt{a_2 n}} \\
    \end{split}
    \label{critical_exps_single_potential}
\end{equation}
controlling the growing of the potential as we approach the codimension-2 intersection between the two codimension-1 branes.

As anticipated, the new key feature of these solutions is that they go beyond the conformally flat ansatz of sections \ref{section:Intersecting_ETW} and \ref{section:decoupled_scalars}. Indeed, using a convenient redefinition of the spacetime coordinates we can write the metric \eqref{Solutions:class2} in the following form:
\begin{equation}
    \begin{split}
         ds^2_{n+2} = & y_1^{2(1-a_1n)}y_2^{2 a_2} \left[ y_1^{4a_1n-2} ds^2_n+y_1^{-2(1-a_1n)} dy_1^2 - y_2^{-2a_2} dy_2^2 \right] = \\
         & \propto \tilde{y}_1^{2 \frac{(1-a_1n)}{a_1 n}} \tilde{y}_2^{\frac{2a_2}{1-a_2n}} \left[ \tilde{y}_1^{4- \frac{2}{a_1 n}} ds^2_n + d \tilde{y}_1^2 - d \tilde{y}_2^2 \right], \\
    \end{split}
\end{equation}
 which is not conformally flat due to the presence of the extra factor in front of the $n$ dimensional metric $ds^2_n$.

 The conformally flat solutions in section \ref{section:Intersecting_ETW} were shown in \cite{Angius:2023rma} to correspond to the backreaction of the superposition of two codimension-1 source terms, with no codimension-2 sources. In our solutions the extra factor beyond the conformally flat ansatz suggests the presence of an additional codimension-2 source localized at the intersection between the two ETW branes.. It would be interesting to explore this further.
 
 \subsubsection{Scaling Relations}

As we point out above, one interesting feature of the Dynamical Cobordism solutions is the existence of specific scaling relations linking the spacetime and the field space distances $\Delta$ and $\mathcal{D}$ with the spacetime scalar curvature $\vert R \vert$. Such relations are summarized in \eqref{scaling_relations_cod1_ETW} for the case of codimension-1 singularities and they are controlled by the critical exponent $\delta$ characterizing the type of ETW brane dressing the singularity. In \cite{Angius:2023xtu} these scaling relations were extended to the case when we approach the codimension-2 singularity located at the intersection between two distinct codimension-1 ETW branes. In these setups, the parameter controlling the scaling is a path-dependent combination of the critical exponents associated with the two individual intersecting branes. In this section we discuss the same relations for the new class of non-conformally flat solutions.  

Consider the following parameterization for the spacetime paths in the region $y_i <1 $ that approach the intersecting locus $y_i=0$ as $t \mapsto 0$:
\begin{equation}
    \Gamma (\gamma_1, \gamma_2) : \quad \begin{cases}
        & y_1 = t^{\gamma_1} \\ & y_2=t^{\gamma_2}
    \end{cases},
    \label{path_parameterization}
\end{equation}
where $\gamma_i$ are two positive real numbers.\\
The spacetime distance to the origin for the path $\Gamma$ identified by the pair of numbers $(\gamma_1 , \gamma_2)$, is:
\begin{equation}
    \Delta = \int_{\Gamma} \big\vert y_2^{2a_2} \left( \partial_t y_1\right)^2 -y_1^{2(1-a_1n)} \left( \partial_t y_2\right)^2 \big\vert^{1/2} dt = \int_{\Gamma} \big\vert \gamma_1^2 t^{2r_1}- \gamma_2^2 t^{2r_2} \big\vert^{1/2},
\end{equation}
with:
\begin{equation}
    \begin{split}
        & r_1 = \gamma_1 -1 +a_2 \gamma_2 \\
        & r_2 = \gamma_2 -1 +(1-a_1 n) \gamma_1. \\
    \end{split}
\end{equation}
The two contributions to the integral are comparable in the case $r_1=r_2$:
\begin{equation}
    \frac{\gamma_1}{\gamma_2} =  \frac{1-a_2}{a_1 n},
\end{equation}
at which the distance $\Delta$ is minimized. In all the other cases one of the two terms dominates over the other and the spacetime distance behaves as:
\begin{equation}
    \Delta = \int_{\Gamma} \gamma_i^2 t^{r_i} dt = \frac{\gamma_i}{r_i +1} t^{r_i +1},
    \label{distance_spacetime}
\end{equation}
with $i=1$ for the paths above the tangent line to the path \eqref{path_parameterization} at $r_1=r_2$, and $i=2$ for the paths below that line. 

Since we know the profiles of the scalars $\phi_i (y_i)$ in terms of the spacetime coordinates, we can translate each spacetime path in \eqref{path_parameterization} into a path in the field space $\phi_i \left( y (t)\right)$ and compute the corresponding distance using the kinetic metric appearing in the spacetime action:
\begin{equation}
    \mathcal{D}  = \int \big\vert d \phi_1^2 +d \phi_2^2 + \alpha d \phi_1 d \phi_2 \big\vert^{1/2} 
\end{equation}
with $\alpha$ given in \eqref{alpha_non_conf_flat}. For each of the two regimes in \eqref{distance_spacetime} we can write the distance $\mathcal{D}$ in the field space in terms of the distance in the spacetime:
\begin{equation}
    \mathcal{D} \sim \frac{\sqrt{n \left[ a_1 \left( 2-a_1-a_1 n \right) \gamma_1^2+2 a_2 (1-a_1 n) \gamma_1 \gamma_2 +a_2 \gamma_2^2 \right]}}{r_i +1} \log \Delta.
\end{equation}
Using the parameterization $\gamma_1 = \sqrt{\gamma}$ and $\gamma_2 = 1/ \sqrt{\gamma}$ such that the ratio $\gamma_1 / \gamma_2 = \gamma$ and the value of $\gamma$ separating the two regimes is $\gamma^{\ast} = (1-a_2)/a_1 n$ we obtain the scaling relation:
\begin{equation}
    \Delta \sim e^{- \frac{1}{2} \delta_{int} \mathcal{D}}
    \label{scaling_relation:cod_2}
\end{equation}
with
\begin{equation}
    \delta_{int} = \begin{cases}
       & \frac{ 2 \left( \frac{a_2}{\sqrt{\gamma}} + \sqrt{\gamma} \right)}{\sqrt{ n \left[ \frac{a_2}{\gamma} +a_1 (2-a_1-a_1 n) \gamma + 2 a_2 (1-a_1 n)\right] }} \quad \quad \text{for} \quad \gamma > \gamma^{\ast},  \\
       & \frac{2 \left( \frac{1}{\sqrt{\gamma}} + \sqrt{\gamma} (1-a_1 n) \right)}{ \sqrt{ n \left[ \frac{a_2}{\gamma} +a_1 (2-a_1-a_1 n) \gamma + 2 a_2 (1-a_1 n)\right]}} \quad \quad \text{for} \quad \gamma < \gamma^{\ast}. \\
    \end{cases}
\end{equation}
This result is different with respect to the one discussed in section 3.3 of \cite{Angius:2023rma} due to the modified relation between the parameter $a_1$ and the critical exponent $\delta_1$ in \eqref{critical_exps_single_potential}. Nevertheless, the generalization to non-conformally flat metrics preserves the fact that one still gets nice scaling relations encoding the fundamental properties defining the realization of a dynamical cobordism in the spacetime.   

Notice that in the limits $\gamma \mapsto 0$ and $\gamma \mapsto \infty$ we recover respectively the critical exponent $\delta_2$ and the combination $\delta_1 /(1-a_1n) $ controlling the asymptotic behavior of the fields \eqref{redefined_fields}.

\section{ETW networks for infinite distance limits in CY moduli space}
\label{section:4ETW_networks}
In this section we are finally ready to explore the network of singular divisors located at infinite distance in the complex structure sector of the Calabi-Yau moduli space using ETW brane solutions in spacetime. 

The key tool in this dictionary between infinite distance limits in moduli space and cobordisms to nothing in spacetime is the translation, for each specific infinite distance limit, of the information encoded in the Hodge-Deligne structure and in the flux with that encoded in the critical exponents for ETW branes. As already anticipated, each ETW brane explores a singular divisor, while intersecting configurations explore intersections of divisors in a different way with respect to the naive expectation.  

\subsection{ETW branes for simple singularities}
\label{section:ETW_brane_singularities}
In this section we exploit the classification of singularities that can occur in a two-moduli family of Calabi-Yau fourfolds with primitive Hodge number $\hat{m} \geq 4$, done in \cite{Grimm:2019ixq} and reviewed in section \ref{section:singularities}, to write for each of these boundaries an effective spacetime action and to build codimension-1 Dynamical Cobordism solutions, characterized by their appropriate critical exponent, which explore the infinite distance limit for the corresponding divergent scalar. 

\subsubsection{Generalities}

We consider the three-dimensional effective action obtained compactifying M-theory on the above mentioned Calabi-Yau fourfold. The resulting theory is completely specified by the K\"ahler potential, which determines the Weil-Petersson metric $G_{WP}$, and the three-dimensional scalar potential induced by turning on four-form fluxes $G_4$ on certain four-cycles of the internal space. 

In the regime we are interested, namely the asymptotic regime near the singular divisor $\Delta_1$ and very far away from any intersection, we can choose the local set of coordinate in $\mathcal{M}_{cs}$ introduced above \eqref{disks_topology}, such that the putative divisor is parameterized by the condition $z_1=0$.   
The leading term of the K\"ahler potential in the strict asymptotic regime near the divisor $\Delta_1$ is computed applying the growth theorem \eqref{growth_theorem} to the holomorphic form $\Omega$:
\begin{equation}
    \mathcal{K}^{cs}_{Sl(2)} \simeq - \log \left[ (s_1)^{d_1} f(\xi)\right],
\end{equation}
where $d_1$ identifies the location of the form in the monodromy filtration $W \left( N_1 \right)$. Note that $d_1$ is exactly the number used in section \ref{section:singularities} to classify the five classes of singularities. 

 Using this result we can compute the following expansion for the Weil-Petersson metric:
 \begin{equation}
  G_{s_1 s_1}= \partial_{t_1} \partial_{\bar{t}_1} \mathcal{K}^{cs} = \frac{1}{4} \frac{d_1}{(s_1)^2} + \frac{\sharp}{(s_1)^3} + ... + \mathcal{O} \left( e^{2 \pi i t_1}\right)   
 \end{equation}
and keep only the leading contribution. Note that the constant coefficient of the leading term is completely determined by the integer $d_1$ which contains the information about the specific type of singularity. Applying this computation to each infinite distance singularity in $\mathcal{M}_{cs}$, we can obtain, as done in \cite{Grimm:2019ixq}, the following few possibilities to construct the kinetic sector of the spacetime action:
\begin{center}
\begin{tabular}{|c|c|c|c|}
\hline
Singularity & $d_1$ & K\"ahler potential & $G_{s_1 s_1}$ \\
\hline
\textbf{II} & $1$ & $K^{cs} \sim - \log \left[(s_1)^1 f (\xi) \right]$ & $ \sim \frac{1}{4} \frac{1}{(s_1)^2}$ \\
\hline
\textbf{III} & $2$ & $K^{cs} \sim - \log \left[(s_1)^2 f (\xi) \right]$& $ \sim \frac{1}{4} \frac{2}{(s_1)^2}$ \\
\hline
\textbf{IV} & $3$ & $K^{cs} \sim - \log \left[(s_1)^3 f (\xi) \right]$ & $ \sim \frac{1}{4} \frac{3}{(s_1)^2}$\\
\hline
\textbf{V} & $4$ & $K^{cs} \sim - \log \left[(s_1)^4 f (\xi) \right]$ & $ \sim \frac{1}{4} \frac{4}{(s_1)^2}$ \\
\hline
\end{tabular}
\end{center}

 The introduction of a primitive $G_4$ four-form flux in some internal cycle of the compactification generates an effective potential \eqref{3d_scalar_potential} for the three dimensional action in the Einstein frame. According to section \ref{subsection:SAR}, we have that in the strict asymptotic regimes of the moduli space the middle cohomology split in eigenspaces of the $sl(2)$ operators $Y_i$, \eqref{H4_decomposition}, and the $G_4$ flux decomposes in this splitting as:
\begin{equation}
    G_4 = \sum_{\mathbf{l} \in \mathcal{E}} G_4^{\mathbf{l}}
    \label{G_4_expansion}
\end{equation}
This sum tells us that approaching the putative singular locus, $G_4$ spreads in the corresponding Hodge-Deligne diamond admitting components only in the admissible sectors $V_{\mathbf{l}}$ of $H_p^4 (Y_4, \mathbb{C})$.\\
All this information allows to use the growth theorem for the Hodge norm \eqref{growth_theorem} to extract the leading complex structure dependence of the potential $V_M$ via:
\begin{equation}
    V_M = \frac{1}{\mathcal{V}^3_4} \vert \vert G_4 \vert \vert^2 \sim \frac{1}{\mathcal{V}^3_4} \vert \vert G_4 \vert \vert^2_{sl(2)} = \sum_{l_1 \in \mathcal{E}} \left( s_1\right)^{l_1-4} \vert \vert \rho_{l_1} (G_4, a) \vert \vert^2
\end{equation}
in the asymptotic regime near the singular divisor $\Delta_1$ parameterized by the divergent coordinate $s_1 \mapsto \infty$. The positive coefficient $\vert \vert \rho_{l_1} (G_4, a) \vert \vert^2$ is constraint by the flux quantization conditions and it carries within it the dependence on the real part of the coordinate $t_1$ as in \eqref{local_coordinates}.  

Since in this section we are considering asymptotic regimes near to a specific singular divisor $\Delta_1$ but very far away from any higher intersection, in the previous formula we have a single divergent scalar $s_1$ and one single component $l_1$ for the vector $\mathbf{l}$. Depending on the value of $l_1$ we can have power divergent potentials, whenever $4 < l_1 \leq 8$, or power vanishing potentials, when $0 \leq l_1 <4$.
The leading contribution of the three-dimensional spacetime action for the scalar $s_1$ is:
\begin{equation}
    S = \int d^3x \sqrt{-g} \left[ \frac{1}{2} R -  \frac{d_1}{4 (s_1)^2} \left( \partial s_1 \right)^2 - \vert \vert \rho_{l_1} \vert \vert^2 s_1^{l_1-4}  \right].
    \label{action_single_sk}
\end{equation}
In order to put it in the form \eqref{action:single_scalar} we redefine the field as:
\begin{equation}
  \phi = \sqrt{\frac{d_1}{2}} \log s_1,  
\end{equation}
and the action \eqref{action_single_sk} becomes:
\begin{equation}
    S = \int d^3x \sqrt{-g} \left[ \frac{1}{2} R - \frac{1}{2} \left( \partial \phi \right)^2 - \vert \vert \rho_{l_1} \vert \vert^2  e^{\sqrt{\frac{2}{d_1}} (l_1-4) \phi} \right]
    \label{action_single_divisor}
\end{equation}
For our analysis, we consider in the above action only power divergent potentials, with $4 < l_1 \leq 8$. This is because for decreasing potentials the Dynamical Cobordism solutions are all equivalent to the case with no potential, and the value of $\delta$ is fixed to a specific number for all the different types of singularities. This excludes the possibility to distinguish between the different cases.

As shown in \cite{Angius:2022aeq} and revisited in section \ref{Cod_1_ETW}, the equations of motion associated with this action admit a special class of solutions \eqref{cod-1_solutions} realizing a codimension-1 Dynamical Cobordism to nothing. It is remarkable that the well-established setup of M-theory flux compactification leads to solutions of this type requiring a non-trivial time dependence. It would be interesting to explore the possible cosmological implications of this result and potential relations with phenomena of nucleation of bubbles of something \cite{Friedrich:2024cob}. The important point is that these solutions provide a way to explore the strict asymptotic regime of the corresponding boundary $\Delta_1$ in $\mathcal{M}_{cs}$ through its spacetime realizations as an ETW brane configuration. From equation \eqref{G_4_expansion} we have that the $G_4$ flux of our compactification can have several components turned on that live in different vector spaces $W_{l_1} \left( N_1 \right)$. The relevant term in this expansion for our analysis is the one that grows faster than the others in the $s_1 \mapsto \infty$ limit. Considering the label $l_1$ associates with this leading term  we compute the critical exponent identifying the specific solution in the family through the formula: 
\begin{equation}
    \delta = \sqrt{\frac{2}{d_1}} \left( l_1 -4 \right).
    \label{delta_singular_divisor}
\end{equation}
Let us emphasize that $\delta$ completely determines the behavior of the spacetime fields because it encodes the information about the growth of the potential as we approach to the ETW boundary. From a geometric point of view, such information is completely encoded in the Hodge-Deligne splitting of the middle cohomology near the corresponding singular divisor plus the information about the flux.

\subsubsection{Examples}
Let us consider the example of the type $II_{0,0}$ singularity in the complex structure moduli space of Calabi-Yau fourfold with $h^{3,1}=2$. Its Hodge-Deligne diamond is depicted in Figure \ref{fig4} and it encodes the information about how the Hodge structure of $H^4_p (Y_4, \mathbb{C})$ is spread in the Hodge-Deligne splitting as we approach the singularity. We have the following non trivial groups:
\begin{equation}
\begin{split}
& (i^{0,3}, i^{1,2}, i^{2,1},i^{3,0}) = (1,0,0,1)\\
& (i^{0,4},i^{1,3},i^{2,2},i^{3,1},i^{4,0}) =(0,1,\widehat{m},1,0)\\
& (i^{1,4},i^{2,3},i^{3,2},i^{4,1})= (1,0,0,1)\\
\end{split}
\end{equation}
which implies three possibilities for the location of $G_4$ in the splitting: $l_1=3,4,5$. We neglect the cases $l_1=4,3$ because, according to \eqref{action_single_divisor}, they produce vanishing potential, while for the case $l_k=5$ the equation \eqref{delta_singular_divisor} gives:
\begin{equation}
\text{Type  } II_{0,0} \quad \quad \text{with  } l_k=5 \quad \quad \longrightarrow \quad \quad \delta = \sqrt{2}.
\end{equation}
Following the same procedure for all the  singularities classified in section \ref{section:singularities}, with a $G_4$ flux specified by its location $l_1$ in the splitting, we are able to associate to each of these possibilities a specific spacetime solution characterized by the corresponding critical exponent. The results are summarized in Table \ref{table:2}
\begin{center}
\begin{table}[h!]
\centering
\begin{tabular}{|c|c|c|c||c|c|c|c|}
\hline
\textbf{Type} & $d_k$ & $l_k$ &  $\delta$ & \textbf{Type} & $d_k$ & $l_k$ &  $\delta$ \\
\hline
$II_{0,0}$ & $1$ & $5$& $ +\sqrt{2}$ & $II_{0,1}$ & $1$ & $5$ & $+\sqrt{2}$ \\
\hline
$II_{1,1}$ & $1$ & $5$& $ +\sqrt{2}$ & $II_{1,1}$ & $1$ & $6$ & $+2\sqrt{2}$ \\
\hline
$III_{0,0}$ & $2$ & $6$& $ +2$ & $III_{1,1}$ & $2$ & $6$ & $+2$ \\
\hline
$III_{0,1}$ & $2$ & $5$& $ +1$ & $III_{0,1}$ & $2$ & $6$ & $+2$ \\
\hline
$IV_{0,1}$ & $3$ & $5$& $ +\sqrt{6}/3$ & $IV_{0,1}$ & $3$ & $7$ & $+\sqrt{6}$ \\
\hline
$V_{1,1}$ & $4$ & $6$& $ +\sqrt{2}$ & $V_{1,1}$ & $4$ & $8$ & $+2\sqrt{2}$ \\
\hline
$V_{1,2}$ & $4$ & $5$& $ +\sqrt{2}/2$ & $V_{1,2}$ & $4$ & $6$ & $+\sqrt{2}$ \\
\hline
$V_{1,2}$ & $4$ & $8$& $ +2\sqrt{2}$ & $V_{2,2}$ & $4$ & $6$ & $+\sqrt{2}$ \\
\hline
$V_{2,2}$ & $4$ & $8$& $ +2\sqrt{2}$ &  &  &  &  \\
\hline
\end{tabular}
\caption{Critical exponents characterizing the ETW brane solutions in spacetime for each type of singularity that can occur in the complex structure moduli space of Calabi-Yau fourfold with $h^{1,3}=2$.}
\label{table:2}
\end{table}
\end{center}
Notice that all the critical exponents classified in the table stay inside the upper bound $\delta \leq 2 \sqrt{\frac{d-1}{d-2}}$, that for three spacetime dimensions means $\delta \leq 2 \sqrt{2}$. Since we are dealing with positive potentials, this implies that the spacetime solutions of the form \eqref{cod-1_solutions} realizing these infinite distance limits involve a timelike running coordinate. The cases $II_{1,1}$ with a non-trivial flux component in $l_1=6$ and $V_{1,1}, V_{1,2}$ and $V_{2,2}$ with $l_1=8$ are special because they saturate the bound for the critical exponent: they correspond in regimes where the scalar potential is subleading with respect to the kinetic term, so they behave as in the case of zero potential.

\subsection{ETW networks for the enhancements}
\label{section:ETW_branes_enhancements}
The main claim of the previous section is the capability to explore any singular divisor of the Calabi-Yau moduli space with an ETW brane solution in spacetime. In this section we propose the use of real codimension-2 intersecting ETW brane solutions to explore the network of intersecting divisors in CY moduli space with flux potential.

As anticipated in section \ref{section:decoupled_scalars}, the naive expectation is that individual ETW branes explore singular divisors in the moduli space, according to the dictionary of the previous section, and their intersection would explore the intersection between the corresponding divisors. However the actual interpretation of the spacetime solutions turns out to be more subtle, and much closer to the mathematical description of the network of divisors in terms of a structure of singularity enhancements in specific growth sectors, as we now explain.

\subsubsection{Generalities}
The regime we are interested is the strict asymptotic regime that approaches the codimension-2 singularity $\Delta_{12}= \Delta_1 \cap \Delta_2$ through the growth sector $R_{12}$, staying very far away from any higher intersection. As explained in section \ref{subsection:AR_NOT}, in this region we can fix a local set of coordinates $\left\lbrace t_1 , t_2, \xi^{\bar{j}} \right\rbrace$, such that the putative locus $\Delta_{12}$ is parameterized by the conditions $s_1 \succ s_2 \mapsto \infty$. Applying the growth theorem \eqref{growth_theorem} to the Hodge norm of the holomorphic form $\Omega$, we can extract the leading term of the K\"ahler potential in the sector $R_{12}$:
\begin{equation}
    \mathcal{K}^{cs}_{Sl(2)} \simeq - \log \left[ (s_1)^{d_1} (s_2)^{(d_e -d_1)} f (\xi) \right],
    \label{Kahler_potential_2moduli}
\end{equation}
where $d_1$ labels the type of singularity associated with the divisor $\Delta_1$, according with the classification of section \ref{section:singularities}, and $d_e$ labels the type of singularity occurring at the intersection $\Delta_{12}$, according with the allowed enhancements listed in the table \ref{table_enhancements}. Using the equation \eqref{Kahler_potential_2moduli}, we compute for all the possible enhancements of $\mathcal{M}_{cs}$, listed in table \ref{table_enhancements}, the corresponding asymptotic Weil-Petersson metric via \eqref{WP_metric} and we summarize the results in table \ref{table:4}.

\begin{table}[h!]
\centering
\begin{tabular}{|c|c|c|c|c|}
\hline
\textbf{Enhancement} & $d_1$ & $d_e$ & \textit{K\"ahler Potential} & \textit{WP Metric}  \\
\hline
$\textit{II}_{0,0} \mapsto \textit{II}_{0,1}$ & $1$ & $1$ & $\mathcal{K}^{cs} \sim - \log \left[ s_1 f (\xi) \right]$ & $G_{s_i s_j} \sim \left( \begin{matrix} \frac{1}{4(s_1)^2} & 0 \\ 0 & 0  \end{matrix} \right)$  \\
\hline
$\textit{II}_{0,0} \mapsto \textit{II}_{1,1}$ & $1$ & $1$ & $\mathcal{K}^{cs} \sim - \log \left[ s_1 f (\xi) \right]$ & $G_{s_i s_j} \sim \left( \begin{matrix} \frac{1}{4(s_1)^2} & 0 \\ 0 & 0 \end{matrix} \right)$  \\
\hline
$\textit{II}_{0,1} \mapsto \textit{II}_{1,1}$ & $1$ & $1$ & $\mathcal{K}^{cs} \sim - \log \left[ s_1 f (\xi) \right]$ & $G_{s_i s_j} \sim \left( \begin{matrix} \frac{1}{4(s_1)^2} & 0 \\ 0 & 0 \end{matrix} \right)$   \\
\hline
$\textit{II}_{0,1} \mapsto \textit{III}_{0,0}$ & $1$ & $2$ & $\mathcal{K}^{cs} \sim - \log \left[ (s_1)(s_2) f (\xi) \right]$ & $G_{s_i s_j} \sim \left( \begin{matrix} \frac{1}{4(s_1)^2} & 0 \\ 0 & \frac{1}{4(s_2)^2} \end{matrix} \right)$   \\
\hline
$\textit{II}_{0,1} \mapsto \textit{V}_{2,2}$ & $1$ & $4$ & $\mathcal{K}^{cs} \sim - \log \left[ (s_1)(s_2)^3 f (\xi) \right]$ & $G_{s_i s_j} \sim \left( \begin{matrix} \frac{1}{4(s_1)^2} & 0 \\ 0 & \frac{3}{4(s_2)^2} \end{matrix} \right)$  \\
\hline
$\textit{III}_{1,1} \mapsto \textit{V}_{2,2}$ & $2$ & $4$ & $\mathcal{K}^{cs} \sim - \log \left[ (s_1)^2 (s_2)^2 f (\xi) \right]$ & $G_{s_i s_j} \sim \left( \begin{matrix} \frac{1}{2(s_1)^2} & 0 \\ 0 & \frac{1}{2(s_2)^2} \end{matrix} \right)$  \\
\hline
$\textit{III}_{0,0} \mapsto \textit{III}_{0,1}$ & $2$ & $2$ & $\mathcal{K}^{cs} \sim - \log \left[ (s_1)^2 f (\xi) \right]$ & $G_{s_i s_j} \sim \left( \begin{matrix} \frac{1}{4(s_1)^2} & 0 \\ 0 & 0 \end{matrix} \right)$ \\
\hline
$\textit{III}_{0,0} \mapsto \textit{III}_{1,1}$ & $2$ & $2$ & $\mathcal{K}^{cs} \sim - \log \left[ (s_1)^2 f (\xi) \right]$ & $G_{s_i s_j} \sim \left( \begin{matrix} \frac{1}{4(s_1)^2} & 0 \\ 0 & 0 \end{matrix} \right)$ \\
\hline
$\textit{III}_{0,1} \mapsto \textit{III}_{1,1}$ & $2$ & $2$ & $\mathcal{K}^{cs} \sim - \log \left[ (s_1)^2 f (\xi) \right]$ & $G_{s_i s_j} \sim \left( \begin{matrix} \frac{1}{4(s_1)^2} & 0 \\ 0 & 0 \end{matrix} \right)$  \\
\hline
$\textit{IV}_{0,1} \mapsto \textit{V}_{2,2}$ & $3$ & $4$ & $\mathcal{K}^{cs} \sim - \log \left[ (s_1)^3 (s_2) f (\xi) \right]$ & $G_{s_i s_j} \sim \left( \begin{matrix} \frac{3}{4(s_1)^2} & 0 \\ 0 & \frac{1}{4(s_2)^2} \end{matrix} \right)$  \\
\hline
$\textit{V}_{1,1} \mapsto \textit{V}_{2,2}$ & $4$ & $4$ & $\mathcal{K}^{cs} \sim - \log \left[ (s_1)^4  f (\xi) \right]$ & $G_{s_i s_j} \sim \left( \begin{matrix} \frac{1}{(s_1)^2} & 0 \\ 0 &0 \end{matrix} \right)$  \\
\hline
$\textit{V}_{1,2} \mapsto \textit{V}_{2,2}$ & $4$ & $4$ & $\mathcal{K}^{cs} \sim - \log \left[ (s_1)^4  f (\xi) \right]$ & $G_{s_i s_j} \sim \left( \begin{matrix} \frac{1}{(s_1)^2} & 0 \\ 0 &0 \end{matrix} \right)$  \\
\hline
$\textit{V}_{1,1} \mapsto \textit{V}_{1,2}$ & $4$ & $4$ & $\mathcal{K}^{cs} \sim - \log \left[ (s_1)^4  f (\xi) \right]$ & $G_{s_i s_j} \sim \left( \begin{matrix} \frac{1}{(s_1)^2} & 0 \\ 0 &0 \end{matrix} \right)$ \\
\hline
\end{tabular}
\caption{Leading behavior of the infinite distances in the moduli space from a double intersection singularity.}
\label{table:4}
\end{table}

Note that, except for the enhancements:
\begin{equation*}
    \begin{split}
         \textit{(i)} \quad \textit{II}_{0,1} \mapsto \textit{III}_{0,0} \quad \quad &  \quad \quad \textit{(ii)} \quad \textit{II}_{0,1} \mapsto \textit{V}_{2,2} \\
         \textit{(iii)} \quad \textit{III}_{1,1} \mapsto \textit{V}_{2,2} \quad \quad & \quad \quad \textit{(iv)} \quad \textit{IV}_{0,1} \mapsto \textit{V}_{2,2} \\
    \end{split}
\end{equation*}
whose corresponding Weil-Petersson metrics show a diagonal form, for the rest of the enhancements we obtain degenerate metrics. This implies that these cases cannot be studied with the methods we developed in section \ref{section3:DC}. Computations of the periods beyond the polynomial approximation (see \cite{Bastian:2021bgh} and  \cite{Bastian:2023bhs}) that include exponential corrections lead to non-degenerate metrics for all the enhancements of table \ref{table:4}.  In \cite{Marchesano:2023fll} it was argued that loci associated with leading order degenerate metrics in the vector multiplet moduli space of type IIA string theory compactified on a Calabi-Yau threefold correspond to EFTs containing a subsector that decouples to gravity at infinite distance. It would be interesting to investigate whether such an upcoming gauge theory allows a deep exploration of the ETW branes associated with these regimes.        

As explained in the previous section, turning on a primitive $G_4$-flux in some internal cycles of the compactification we generate an effective three dimensional potential for the Calabi-Yau moduli according with the definition \eqref{3d_scalar_potential}. The $sl(2)$ decomposition \eqref{H4_decomposition} of the middle cohomology near the asymptotic locus $\Delta_{12}$ allows to write the $G_4-$flux as in \eqref{G_4_expansion}, where the allowed doublets $\mathbf{l}= \left(l_1, l_e \right)$ appearing in the expansion are the one classified in table \ref{table_enhancements}. Using these allowed doublets in the growth theorem for the Hodge norm, the authors of \cite{Grimm:2019ixq} compute the complex structure dependence of the flux potential for all the allowed enhancements in $\mathcal{M}_{cs}$ with $h^{3,1}=2$ through the formula:
\begin{equation}
    V_M = \frac{1}{\mathcal{V}^3_4} \vert \vert G_4 \vert \vert^2 \sim \frac{1}{\mathcal{V}^3_4} \vert \vert G_4 \vert \vert^2_{sl(2)} = \sum_{(l_1, l_e) \in \mathcal{E}} \left( s_1\right)^{l_1-4} \left(s_2 \right)^{l_e-l_1} \vert \vert \rho_{(l_1,l_e)} (G_4, a) \vert \vert^2
    \label{V_M:2moduli}
\end{equation}
Since Dynamical Cobordism solutions are not able to distinguish decreasing potentials from the case with zero potential, we consider only the doublets $(l_1,l_e)$ leading to divergent flux-potentials.  In the table \ref{table:5} we summarize the results only for the four enhancements equipped with a non-degenerate metric.
\begin{table}[H]
\centering
\begin{tabular}{|c|c|c|}
\hline
\textbf{Enhancement} & \textit{Doublets} & \textit{Potential} \\
\hline
$\textit{II}_{0,1} \mapsto \textit{III}_{0,0}$ &$(5,6)$, $(5,4)$ & $\sim c_1 (s_1)(s_2) +c_2 \frac{s_1}{s_2}$ \\
\hline
$\textit{II}_{0,1} \mapsto \textit{V}_{2,2}$ & $(5,8)$, $(5,6)$, $(5,4)$, $(5,2)$ &  $\sim c_1 (s_1)(s_2)^3 +c_2 (s_1)(s_2)+c_3 \frac{s_1}{s_2} + c_4 \frac{s_1}{(s_2)^3} $ \\
\hline
$\textit{III}_{1,1} \mapsto \textit{V}_{2,2}$ & $(6,8)$, $(6,6)$, $(6,4)$ &  $\sim c_1 (s_1)^2(s_2)^2+c_2 (s_1)^2+c_3 (s_2)^2 $ \\
\hline
$\textit{IV}_{0,1} \mapsto \textit{V}_{2,2}$ & $(7,8)$, $(7,6)$, $(5,6)$, $(5,4)$ & $\sim c_1 (s_1)^3(s_2)+c_2 \frac{(s_1)^3}{(s_2)}+c_3 (s_1) (s_2) +c_4 \frac{s_1}{s_2}  $ \\
\hline
\end{tabular}
\caption{Divergent potentials associated to the enhancements in $\mathcal{M}_{cs}$ with $h^{3,1}=2$ equipped with a non-degenerate metric. The parameters $c_i$ are positive constant coefficients constrained by the flux quantization conditions.}
\label{table:5}
\end{table}

\subsubsection{Spacetime solutions and interpretation}

Using the Weil-Petersson metric and the dominant contribution of the scalar potential for all the possible enhancements reviewed in the previous section and all the possible components for the $G_4-$flux, we can write the leading contribution of the spacetime action for the involved moduli, valid in the growth sector $R_{12}$, near the intersection locus $\Delta_{12} =\Delta_1 \cap \Delta_2$:
\begin{equation}
    S = \int d^3x \sqrt{-g} \left[ \frac{1}{2} R - \frac{d_1}{4(s_1)^2} \left( \partial s_1 \right)^2 - \frac{d_e-d_1}{4(s_2)^2} \left( \partial s_2 \right)^2 - \vert \vert \rho_{\mathbf{l}} \vert \vert^2 s_1^{l_1-4} s_2^{l_e-l_1} \right].
    \label{action_3d_intersection}
\end{equation}

In order to put the action in the form \eqref{action_2decoupled_scalars}, we  fix the canonical normalization for the scalars through the following redefinition of the fields:
\begin{equation}
    \psi_1 = \sqrt{\frac{d_1}{2}} \log s_1 \quad \quad , \quad \quad \psi_2 = \sqrt{\frac{d_e-d_1}{2}} \log s_2,
\end{equation}
and the action \eqref{action_3d_intersection} takes the form:
\begin{equation}
    S = \int d^3x \sqrt{-g} \left[ \frac{1}{2} R - \frac{1}{2} \left( \partial \psi_1 \right)^2 - \frac{1}{2} \left( \partial \psi_2 \right)^2 - \vert \vert \rho_{\mathbf{l}} \vert \vert^2 e^{(l_1-4) \sqrt{\frac{2}{d_1}} \psi_1} e^{(l_e-l_1) \sqrt{\frac{2}{d_e-d_1}} \psi_2}\right].
    \label{action_CY_2moduli}
\end{equation}
As pointed out above, the diagonal kinetic metric appearing in this spacetime action prevents the naive interpretation of the two scalars as two independent moduli. This is due to the fact that intersecting ETW brane configurations contain a mixed kinetic term in their spacetime action \eqref{action_2scalars} which is always non-zero, \eqref{alpha}, as imposed by the equations of motion. The key idea to solve this drawback is encoded in the redefinition of the fields \eqref{field_redefinition} and \eqref{redefined_fields} relating the decoupled scalars $\psi_i$ with the scalars $\phi_i$ satisfying the equations of motion associated to the action \eqref{action_2scalars} with a specific non-trivial mixing term as in section \ref{section:Intersecting_ETW}. The solutions presented in section \ref{section:decoupled_scalars} for decoupled scalars have the nice property that when we approach the locus $y_1=0$ only the scalar $\psi_1$ diverges, while when we approach the locus $y_2=0$ both the scalars $\psi_1$ and $\psi_2$ diverge at the same time. On the other hand, the solutions with respect to the fields $\phi_i$ have a clear interpretation for these loci as two intersecting ETW branes. In terms of Calabi-Yau moduli, the ETW brane located at the position $y_1=0$, at which the modulus $\psi_1$ diverge, is interpreted as the spacetime realization of the divisor $\Delta_1$ involved in the intersection $\Delta_{12}$. The brane located at position $y_2=0$, at which both the Calabi-Yau moduli diverge, is then naturally interpreted as the spacetime realization of the intersection $\Delta_{12}$ of the two divisors, i.e. the enhancement. This new interpretation matches very well with the description of the network of divisors provided in section \ref{section2:generalities_CY_moduli} where the asymptotic Hodge theory strongly constrains how to jump from a singular divisor $\Delta_1$ to the enhanced singularity at the intersection $\Delta_{12}$ through the growth sector $R_{12}$. Here we have something very similar in the spacetime: the solution describes in the same configuration what happens near the singular divisor $\Delta_1$, corresponding to the ETW brane located at $y_1=0$, and near its enhanced singularity in $\Delta_{12}$, corresponding to the ETW brane located at $y_2=0$.  

Although the interpretation of the intersecting ETW configurations as spacetime realizations of the enhancements in the moduli space along specific growth sectors nicely solve the problem to have asymptotic diagonal metrics, the structure of the potential that supports the class of solutions in section \ref{section:decoupled_scalars} is not of the same form of the potential that we have in the Calabi-Yau moduli space near the enhanced singularities. The potential \eqref{potential:solution1} contains two different contributions: the first one dominating in a cone region near the ETW brane located at the position $y_1=0$, and the second one dominating in the complementary region near the ETW brane located at $y_2=0$. On the other hand, in the spacetime action for the Calabi-Yau moduli the potential is given by a single term: the leading contribution in the sum \eqref{V_M:2moduli} when we approach the intersection between the two different divisors. Then, there is no straight direct way to match the two expressions.

One possibility is to focus on trajectories in the moduli space where one term of the potential \eqref{potential:solution1} is subleading with respect to the other and match with \eqref{V_M:2moduli} only the dominant term of \eqref{potential:solution1}. However, this approach does not provide a solution in the full spacetime region between the two intersecting ETW branes, but only in the sector in which the trajectories selecting the dominant term in the potential are supported.       

Happily, a fully satisfactory realization can be provided using the new class of solutions described in section \ref{section:Non-conformal-solutions}, driven by a single potential term \eqref{potential:solution2}, and performing a special redefinition of the fields \eqref{redefined_fields} which leads to the interpretation of these multiple infinite distance limits as configurations of intersecting ETW branes. 

Note that for each particular enhancement $\Delta_1 \rightarrow \Delta_{12}$, characterized by its corresponding variation of the mixed Hodge-Deligne structure and equipped with the flux information, we have a specific spacetime action \eqref{action_CY_2moduli} defined in terms of the pairs of parameters $\left(d_1, d_e \right)$ and $\left(l_1, l_e \right)$. The main advantage of the solutions computed in section \ref{section:Non-conformal-solutions}, whose potential \eqref{potential:solution2} matches perfectly the structure of the Calabi-Yau flux potential given by the $sl(2)-$approximation of the Hodge norm of $G_4$, is that we can associate to each of these actions a specific solution of the form \eqref{Solutions:class2} and interpret each growth sector $R_{12}$ as an intersecting configuration of branes in spacetime.

\subsubsection{Examples}

In this section we will perform the explicit matching between the potential \eqref{V_M:2moduli}, appearing in the action \eqref{action_CY_2moduli}, and the Dynamical Cobordism potential \eqref{potential:solution2}, with $n=1$, supporting the class of solutions \eqref{Solutions:class2}, for each enhancement of table \ref{table:5} and for each possible leading term of the corresponding potential. The matching produces for each of these enhancements the exact values of the parameters $a_i$, which uniquely determine the spacetime solutions via \eqref{Solutions:class2}. The list of values of $a_i$ obtained for all the allowed enhancements of table \ref{table:5} which are equipped with a non-trivial diagonal kinetic metric, computed in table \ref{table:4}, and with a divergent flux potential showing an explicit dependence from both the moduli is summarized in table \ref{table:3new}. 

\begin{table}[h!]
\centering
\begin{tabular}{|c|c|c|c|c|c|}
\hline
Enhancement & $(l_2,l_e)$ & $a_1$ & $a_2$ & $\delta_1$ & $\delta_2$ \\
\hline
$II_{0,1} \mapsto III_{0,0}$ & $(5,6)$ & $1/2$ & $1$ & $ \sqrt{2}$ & $2$ \\
\hline
$II_{0,1} \mapsto III_{0,0}$ & $(5,4)$ & $1/2$ & $1$ & $ \sqrt{2}$ & $2$ \\
\hline
$II_{0,1} \mapsto V_{2,2}$ & $(5,8)$ & $1/2$ & $1/2$ & $\sqrt{2}$ & $2 \sqrt{2}$ \\
\hline
$II_{0,1} \mapsto V_{2,2}$ & $(5,6)$ & $1/2$ & $3/2$ & $ \sqrt{2}$ & $2 \sqrt{6}/3$ \\
\hline
$II_{0,1} \mapsto V_{2,2}$ & $(5,4)$ & $1/2$ & $3/2$ & $ \sqrt{2}$ & $2 \sqrt{6}/3$ \\
\hline
$II_{0,1} \mapsto V_{2,2}$ & $(5,2)$ & $1/2$ & $1/2$ & $\sqrt{2}$ & $2 \sqrt{2}$ \\
\hline
$III_{1,1} \mapsto V_{2,2}$ & $(6,8)$ & $1/3$ & $1/4$ & $2$ & $4 $ \\
\hline
$III_{1,1} \mapsto V_{2,2}$ & $(6,6)$ & $1/3$ & $1$ & $2$ & $2$ \\
\hline
$III_{1,1} \mapsto V_{2,2}$ & $(6,4)$ & $1/2$ & $1/4$ & $2$ & $4$ \\
\hline
$IV_{0,1} \mapsto V_{2,2}$ & $(7,8)$ & $1/4$ & $1/2$ & $ \sqrt{6}$ & $2 \sqrt{2}$ \\
\hline
$IV_{0,1} \mapsto V_{2,2}$ & $(7,6)$ & $1/4$ & $1/2$ & $ \sqrt{6}$ & $2 \sqrt{2}$ \\
\hline
$IV_{0,1} \mapsto V_{2,2}$ & $(5,6)$ & $3/4$ & $3/2$ & $ \sqrt{6} /3$ & $2 \sqrt{6}/3$ \\
\hline
$IV_{0,1} \mapsto V_{2,2}$ & $(5,4)$ & $3/4$ & $3/2$ & $ \sqrt{6}$ & $2 \sqrt{6}/3$ \\
\hline
\end{tabular}
\caption{Table showing the parameters $a_i$ appearing in the spacetime solutions \eqref{Solutions:class2} for each possible enhancement and the corresponding values of the critical exponents characterizing the nature of the intersecting ETW branes.}
\label{table:3new}
\end{table}

In order to obtain the correct sign in the potential we are requiring the direction $y_2$ to be timelike. The corresponding solutions physically describe a configuration of intersecting branes with a boundary in the spacelike direction $y_1$ and an origin in time at the location $y_2=0$. It would be interesting to explore the role of such solutions in cosmological applications. A special configuration with a boundary defining the beginning of time and a boundary with respect to a spacelike coordinate has been explored in \cite{Angius:2022mgh} in a supercritical bosonic string theory setup with tachyon condensation along a lightlike direction. However, in that case there is an interpretation of the two boundaries as two different phases of a same recombinating ETW brane. In the present case of Calabi-Yau moduli we have an interpretation of the two boundaries in terms of an intersecting configuration of two distinct ETW branes.     

Using the definition of critical exponents given in \eqref{critical_exps_single_potential} we can characterize the two intersecting branes appearing in each of these spacetime realizations in the language of the Dynamical Cobordism. The values of the critical exponents for all the accounted enhancements are summarized in the last two columns of table \ref{table:3new}. Notice that the list of values obtained for $\delta_1$ nicely agree with the dictionary between singular divisors and codimension 1 ETW branes provided in the table \ref{table:2} of the previous subsection. As explained above, the ETW brane at $y_2=0$ is interpreted as the spacetime representation of the enhanced singularity occurring at the intersection locus $\Delta_{12}$ in $\mathcal{M}_{cs}$. For that reason the corresponding critical exponent $\delta_2$ does not have to match with the values appearing in table \ref{table:2}: the corresponding ETW brane does not represent a singular divisor in the moduli space but an enhanced singularity.  

\subsection{Application to Swampland Distance Conjecture}
One of the most explored swampland conjecture is the Distance Conjecture \cite{Ooguri:2006in}, which predicts for any infinite distance limit in moduli space the emergence of an infinite tower of states becoming exponentially massless with the field space distance, therefore the cutoff of the corresponding effective theory is lowered as:
\begin{equation}
    \Lambda \sim e^{- \lambda \mathcal{D}}
    \label{distance_conj}
\end{equation}
where $\lambda$ is an $\mathcal{O}(1)$ parameter.

On the other hand, Dynamical Cobordism solutions probes infinite distance limits in field spaces, then it is natural to ask for their interplay. Since the Dynamical Cobordism analysis provides scaling relations linking the spacetime and the field space distances, we can offer a spacetime version of \eqref{distance_conj} expressing the lowering of the cutoff in terms of paths in the spacetime:
\begin{equation}
    \Lambda \sim \Delta^{\frac{2 \lambda}{\delta_{int}}}
    \label{cutoff_EFT}
\end{equation}
This spacetime rephrasing of the Distance Conjecture was proposed in \cite{Angius:2022aeq} for codimension 1 infinite distance limits explored with codimension-1 ETW brane. In \cite{Angius:2023rma}, it was shown that the same relation is still valid for codimension-2 infinite distance limits in the field space probed in spacetime through configurations of intersecting ETW branes. The only difference in this second case, with respect to the case of single ETW brane, is that the relation \eqref{cutoff_EFT} is controlled by a path-dependent parameter $\delta_{int}$ which takes different values for each specific spacetime direction traveled to reach the intersection between the two ETW branes. Due to the knowledge of the profiles of the divergent scalars that parameterize the codimension-2 infinite distance locus in the field space in terms of the spacetime coordinates, it is possible to map each of these spacetime paths to a specific path in field space that reaches the multiple infinite distance limit, and to obtain the decay rate of $\Lambda$ for each of these paths.  

For the class of solutions reviewed in section \ref{section:Intersecting_ETW}, considering all the spacetime paths reaching the brane intersection along all the possible different directions, we are able to reproduce all the possible paths in the field space reaching the corresponding codimension-2 infinite distance limit. However, this is not true for the new class of solutions studied in section \ref{section:Non-conformal-solutions}. The new insight lies in the interpretation of these solutions as the intersection of two ETW branes, where one brane represents a codimension-1 infinite distance locus in field space, corresponding to the singular divisor $\Delta_1$, and the other intersecting brane does not represent one of its intersecting divisors $\Delta_2$, but the enhanced singularity arising at the codimension-2 locus $\Delta_{12}= \Delta_1 \cap \Delta_2$. This new interpretation implies that 
none of the paths parameterized in \eqref{path_parameterization} corresponds to a path in field space able to reach the divisor $\Delta_2$. Roughly speaking, using the profiles of the fields given in \eqref{Solutions:class2} to translate the spacetime paths parameterized in \eqref{path_parameterization} to the corresponding ones in the field space, we are able to reproduce only the field space sector containing paths that achieve the locus $\Delta_{12}$ passing close to the divisor $\Delta_1$. In particular, the field space path $\tilde{\Gamma} (\gamma_1 , \gamma_2)$, corresponding to the path $\Gamma (\gamma_1, \gamma_2)$ in spacetime, is:
\begin{equation}
    \tilde{\Gamma} (\gamma_1, \gamma_2) : \quad \begin{cases}
        & \psi_1 = - \left[ \gamma_1 \sqrt{a_1 n (2-a_1-a_1n)} + \gamma_2 a_2 (1-a_1n) \sqrt{\frac{n}{a_1 (2-a_1-a_1n)}} \right] \log t \\
        & \psi_2 = - \gamma_2 \sqrt{a_2 n \left[ 1 + \frac{a_2 (a_1 n-1)^2}{a_1 (-2+a_1-a_1n)} \right]} \log t \\
    \end{cases}
    \label{field_parameterization}
\end{equation}
Since the choice $\gamma_1=0$ in \eqref{path_parameterization} corresponds to approach the ETW brane located at $y_2=0$ following a transversal path to the brane, this direction corresponds to the furthest path from $\Delta_1$ that we can have in the field space. Therefore the parameterization \eqref{field_parameterization} includes all the field space paths that satisfy the condition:
\begin{equation}
    \frac{\psi_1}{\psi_2} > \frac{\sqrt{a_2} (1-a_1n)}{\sqrt{a_1(-2+a_1+a_1n) +a_2(a_1n-1)^2}}.
\end{equation}
This interpretation nicely reproduces the growth sectors $R_{12}$ of the Calabi-Yau moduli space.

\section{Conclusions}
\label{section:5_conclusions}
M-theory compactifications on Calabi-Yau fourfolds with four-form flux produce a rich landscape of vacua reproducing interesting phenomenology. The infinite distance limits of the corresponding moduli/field spaces offer an intriguing arena to test many of the Swampland conjectures and get information about the mechanism breaking the effective field theory description by quantum gravity effect. Interestingly, the structure of the infinite distance limits in the moduli space displays a rich network of intersecting divisors, characterized by the rich mathematical tools of asymptotic Hodge theory \cite{Schmid:1973,Cattani:1982cka,kerr2019polarized,Grimm:2019ixq,Grimm:2018cpv, Grimm:2018ohb, Grimm:2019grh, Bastian:2020bgh}. In this paper we have initiated the study of this network of normal crossing divisors using Dynamical Cobordism solutions in the resulting three dimensional effective field theories. Some of our results in this context are: 
\begin{itemize}
    \item We have provided a dictionary associating to each singular divisor in the network its spacetime realization as a codimension-1 ETW brane characterized by a specific critical exponent; 
    \item We have shown that, due to the properties of the flux potential, this match requires the corresponding Dynamical Cobordism to describe a time-dependent solution; 
    \item We have extended this dictionary to codimension-2 loci in the network of divisors at infinity in the Calabi-Yau moduli space by using codimension-2 Dynamical Cobordisms describing intersecting ETW branes; 
    \item We have shown that the two intersecting ETW branes in the spacetime picture, rather than describe two intersecting divisors in the moduli space, describe the singularity enhancement of a divisor as it approaches a singularity in a specific growth sector. It is remarkable that the spacetime picture reproduces the spirit of the mathematical approach to the study of intersections of divisors; 
    \item We have shown that, in order to match the leading behaviour of the flux potential given by the asymptotic growth of the Hodge norm, the required spacetime solutions for intersecting ETW branes  are more general than those considered hitherto and we have provided the explicit construction of such generalization, by relaxing the constraint of conformally flat ansatz in the solutions in \cite{Angius:2023rma};  
    \item  We studied the scaling relations between the spacetime distance and the field space distance along general paths in the new intersecting ETW brane solution, generalizing those in the literature, and  we explored its relation to the cutoff implied by the Swampland Distance Conjecture, thus providing a generalization valid for non-trivial scalar potentials in Calabi-Yau flux compactifications.
    \end{itemize}
    
Some related interesting open questions are the following:
\begin{itemize}
    \item It would be interesting to further study the class of non-conformally flat solutions involving two divergent scalar fields we constructed in this paper. For instance to display the source terms supporting it and to characterize the properties of the possible codimension-2 source term localized at the intersection of the ETW branes, and study possible connections with other setups including such codimension-2 sources \cite{Blumenhagen:2022mqw}.   
    \item We performed explicit computations of the critical exponents characterizing the ETW branes associated to each possible singular divisor and some of the enhanced singularities in the complex structure moduli space of Calabi-Yau fourfolds with $h^{1,3}=2$. It would be interesting to reproduce the same analysis in higher dimensional moduli space, thus allowing for the exploration of higher codimension intersections in the Calabi-Yau moduli space.
    \item In our dictionary between singular divisors in the moduli space and ETW branes in spacetime, we translate the information about the Hodge-Deligne splittings characterizing the moduli space singularities in terms of the critical exponents characterizing the branes. However the inverse is not true: the known class of Dynamical Cobordism solutions does not reproduce all the singularity enhancements in the Calabi-Yau moduli space. In particular, those involving degenerate metrics at infinity in field space may require, from the Calabi-Yau side, a better characterization of the subleading contributions to the Calabi-Yau moduli metric, and from the Dynamical Cobordism side, a generalization of the solutions for scalars with non-trivial metrics for the running scalars. It would be interesting to fully understand the spacetime realizations of the enhancements in order to investigate if the complete picture contains more details allowing a 1-1 correspondence between the two approaches.   
    
\end{itemize}
Our work has provided an important step in the exploration of the intricate structure of the asymptotic Calabi-Yau moduli space in the presence of general flux potential. We hope to come back to these exciting open questions in the future.

\section*{Acknowledgments}
We would like to acknowledge Angel Uranga for his fundamental contribution to this manuscript through meticulous comments and extensive discussions. We are also pleased to thank José Calderon-Infante, Matilda Delgado, Jesús Huertas and Andriana Makridou for useful discussions and for collaborations on related topics. We also thank Ivano Basile, Stefano Giaccari, Alvaro Herráez, Elias Kiritsis, Fernando Marchesano, Anthony Massidda, Irene Valenzuela and Roberto Volpato for helpful discussions. R.A. wishes to acknowledge the hospitality of the Max Planck Institute for Physics in Munich during the last stages of this work. The research of R.A. has been supported by the grants CEX2020001007-S and PID2021-123017NB-I00, funded by MCIN/AEI/10.13039/501100011033 and by ERDF A way of making Europe. 
\newpage
\bibliographystyle{JHEP}
\bibliography{mybib}
\end{document}